\documentclass[a4paper,12pt]{article}
\usepackage{amsmath}
\usepackage{amssymb}
\usepackage{arrayjobx}
\numberwithin{equation}{section}
\usepackage[nosort]{cite}
\usepackage{subfigure}
\usepackage{graphicx}
\usepackage{color} 
\usepackage{bbm}
\usepackage{mathrsfs}
\usepackage[bookmarks=true,hyperfigures=true]{hyperref}
\usepackage[utf8]{inputenc}

\usepackage[a4paper,text={470pt,700pt},centering]{geometry}

\hypersetup{
	colorlinks=true,
	citecolor=blue,
	linkcolor=blue,
	filecolor=magenta,      
	urlcolor=blue,
	pdfpagemode=FullScreen,
}

\newcommand{\nn}{\nonumber}
\newcommand{\gm}{{g}_m}
\newcommand{\Bm}{{B}}

\renewcommand{\[}{\begin{equation}\begin{aligned}}
\renewcommand{\]}{\end{aligned}\end{equation}}
\usepackage{braket}

\newcommand{\cl}[1]{\mathcal{#1}}
\newcommand{\pd}{\partial}
\renewcommand{\Im}{\operatorname{Im}}
\renewcommand{\Re}{\operatorname{Re}}
\begin{document}
\thispagestyle{empty}

\newcommand{\tmcl}[1]{\textcolor{red}{#1}}
\begin{flushright}\footnotesize
	\texttt{
%		TCDMATH 
	}
\end{flushright}
\vspace{1cm}

\begin{center}%
	{\LARGE{ Celestial amplitudes on electromagnetic backgrounds:\\ \vspace{0.3cm}T-duality from S-duality
		}\par}%
	
	\vspace{15mm}

   \centerline{ 
   {Tristan McLoughlin}${}^a$,
   {Nathan Moynihan}${}^b$,
   {Andrea Puhm}${}^c{}$
   }

   \bigskip\bigskip

 \noindent{\em${}^a$
School of Mathematics and Hamilton Mathematics Institute, Trinity College, \\Dublin 2, Ireland
}
\bigskip

\noindent{\em${}^b$
Centre for Theoretical Physics, Department of Physics and Astronomy,
Queen Mary University of London, 327 Mile End Road, London E1 4NS, UK
}
\bigskip

\noindent{\em${}^c$
Institute for Theoretical Physics, University of Amsterdam, PO Box 94485, 1090 GL Amsterdam, The Netherlands
}
	
	%%%%%%%%
	\par\vspace{14mm}
	
	\textbf{Abstract} \vspace{5mm}
	
	\begin{minipage}{14cm}
		\parindent=2em

What is the boundary holographic dual of S-duality for gauge theories in asymptotically flat space-times? Celestial amplitudes, by virtue of exhibiting holographic properties of the S-matrix, appear well-suited for studying this question. We scatter electrically and magnetically charged massless scalars off non-trivial electromagnetic potentials such as shockwaves, spin-one conformal primary waves, conformally soft modes and their magnetic duals which we construct. This reveals an intricate relation between conformally soft solutions, descendant CFT three-point functions and, by means of the two-dimensional shadow transform, CFT two-point functions.
By comparing celestial amplitudes on electric and magnetic dual backgrounds, we provide evidence that the four-dimensional flat-space holographic dual of S-duality in Abelian gauge theory is two-dimensional T-duality. 
Moreover, we demonstrate that for two-dimensional boundary actions describing low-energy sectors of the bulk gauge theory, S-duality can be explicitly implemented as a T-duality transformation. 
We show that the Dirac quantisation condition guarantees gauge invariance in the eikonal re-summation for scattering from potentials which for magnetic scalars can be expressed in terms of 't~Hooft loops.

	\end{minipage}
	
\end{center}

\newpage
\tableofcontents

%%%%%%%%%%%%%%%%%%%%%%%%%%%%%%%%%%%%%%%%%%%%%%%%%%%%%%%%%%%%%%%%%%%%%%%%%
%%%%%%%%%%%%%%%%%%%%%%%%%%%%%%%%%%%%%%%%%%%%%%%%%%%%%%%%%%%%%%%%%%%%%%%%%
\section{Introduction}
%%%%%%%%%%%%%%%%%%%%%%%%%%%%%%%%%%%%%%%%%%%%%%%%%%%%%%%%%%%%%%%%%%%%%%%%%
%%%%%%%%%%%%%%%%%%%%%%%%%%%%%%%%%%%%%%%%%%%%%%%%%%%%%%%%%%%%%%%%%%%%%%%%%
Despite the inherent non-linearity of non-Abelian gauge theory and General Relativity, the structures that emerge ``on-shell" - colour-kinematics duality, unitarity, positive geometries, twistor theory, soft theorems - have enabled the explicit calculation of S-matrices in Minkowski space-time  to very high orders in loops and legs, see \cite{Badger:2023eqz, Travaglini:2022uwo, Cheung:2017pzi} for recent reviews. However, the computation of scattering amplitudes on non-trivial backgrounds is less well developed and there are conceptual difficulties with even defining the S-matrix in a general curved background. 

Nonetheless, where the space-time is asymptotically flat (AF) the S-matrix is a natural observable and so has been studied in a number of contexts. 
For example, the scattering of radiation from black holes is a problem of long-standing interest and has been approached with a range of methods including using QFT amplitudes, see for example \cite{Sanchez:1976fcl, DeLogi:1977dp, Futterman:1988ni, Glampedakis:2001cx, Bautista:2021wfy}. More recently, tree-level graviton scattering in plane wave space-times has been studied \cite{Adamo:2017nia} and all-multiplicity expressions for tree-level graviton scattering in self-dual radiative space-times were found using twistor methods \cite{Adamo:2022mev}, also \cite{Brown:2023zxm}. In related work, gluon scattering from a self-dual dyon was computed in \cite{Adamo:2024xpc} and for the self-dual Taub-NUT geometry the scattering problem was studied in \cite{Adamo:2023fbj}.

Scattering on non-trivial backgrounds has also received attention in attempts to develop a theory of flat-space holography. 
A key element of the celestial holography proposal is the identification of Mellin transformed scattering amplitudes, also known as celestial amplitudes, with correlation functions of operators in the dual celestial conformal field theory (CCFT) \cite{deBoer:2003vf, He:2015zea, Cheung:2016iub, Pasterski:2016qvg, Pasterski:2017kqt}. This identification is natural, as on-shell momenta for massless particles label points on the two-sphere at the null boundary of the AF spacetime where the Lorentz group acts as the Euclidean global conformal group. The Mellin transform corresponds to a change of basis for asymptotic states from momentum to boost eigenstates, with the boost eigenvalue being identified with the conformal scaling dimension of the dual operator insertion. In large part, this duality has mostly been studied in the context of amplitudes in Minkowski space-time (see \cite{Strominger:2017zoo, Raclariu:2021zjz, Pasterski:2021raf, Pasterski:2021rjz, McLoughlin:2022ljp, Donnay:2023mrd} for overviews and further references), though celestial amplitudes for non-trivial backgrounds have been analysed in several contexts \cite{Costello:2022wso, Fan:2022vbz, deGioia:2022fcn, Gonzo:2022tjm, Costello:2022jpg,Stieberger:2022zyk, Melton:2022fsf, Crawley:2023brz,  Banerjee:2023rni, Ball:2023ukj, Melton:2023lnz}.

One aim of this work is to expand the celestial holography dictionary by extending the computations of \cite{Gonzo:2022tjm} for the scattering of electrically and magnetically charged scalars from potentials. 
This is somewhat simpler than the gravitational case but involves many of the same technical and conceptual issues. Further, by means of the classical double-copy \cite{Monteiro:2014cda}, many computations can be explicitly recycled into the gravitational case. As in \cite{Gonzo:2022tjm}, we consider amplitudes in the leading Born approximation using the Boulware-Brown \cite{Boulware:1968zz} result that the solution to the field equations, in the presence of an arbitrary external
 source function, is the generating functional for the
 connected tree-level Green's functions. This method is very similar to the so-called perturbiner or AFS approach \cite{Arefeva:1974jv, Abbott:1983zw, Jevicki:1987ax} which has also been applied to the perturbative computation of scattering amplitudes in non-trivial backgrounds and related questions e.g. \cite{Adamo:2023cfp, Ilderton:2023ifn, Jain:2023fxc, Kim:2023qbl}.  
 
In \cite{Gonzo:2022tjm} the $1\to1$ scattering amplitude, that is the two-point function, for an electrically charged scalar is computed in the Coulomb and electromagnetic shockwave potentials, as well as for their spinning analogues. One notable feature of the two-point function on shockwave backgrounds is that it can be recast as a standard CFT three-point function. This is due to the fact that the shockwave can be interpreted as a conformal primary \cite{Pasterski:2020pdk}. Additionally, the amplitudes on these backgrounds are less singular than flat space amplitudes in having no kinematic Dirac-deltas and, in the spinning case, improved UV behaviour. In section \ref{sec:bgscat}, we extend the considerations of \cite{Gonzo:2022tjm} to the case of backgrounds corresponding to elements of the basis of massless scalar conformal primary wavefunctions (CPW) introduced in \cite{Pasterski:2017kqt}. These are vacuum solutions of the wave equation that transform as conformal primaries under the Lorentz group and are labelled by a conformal dimension $\Delta$ and a spin $J$. For the principal continuous series with $\Delta\in 1+i \mathbbm{R}$ these wavefunction provide a basis of normalisable solutions\footnote{It has been recently shown that there exists a complete, orthogonal, discrete basis for $\Delta \in \mathbbm{Z}$ \cite{Freidel:2022skz, Cotler:2023qwh}.}. We consider the electrically charged scalar two-point function in the background of massless spin-one CPW and find that due to kinematic constraints the amplitudes vanish for real momenta. For complex momenta, one can solve the kinematic constraints for either holomorphic or anti-holomorphic coordinates and so find non-trivial chiral amplitudes. 

For $\Delta\in \tfrac{1}{2} \mathbbm{Z}$ the primary operators generate asymptotic symmetries and are related to the conformally soft limit of amplitudes \cite{Donnay:2018neh, Pasterski:2020pdk}. 
The $\Delta=1$ spin-1 CPW is the total derivative Goldstone mode related to the asymptotic U$(1)$ Kac-Moody symmetry. The canonical partner of this Goldstone mode is a conformally soft Memory mode which in \cite{Donnay:2018neh} was constructed as a particular linear combination of $(\Delta,J)=(1,\pm1)$ logarithmic solutions. We compute the two-point function in this background and find that the amplitude has a form closely related to the current-scalar-scalar three-point function found in \cite{Chang:2022seh} which used a logarithmic background. 
Upon closer inspection, the logarithmic as well as the conformally soft backgrounds can be expressed as descendants of {\it generalised} conformal primaries as defined in \cite{Pasterski:2020pdk} with $(\Delta',J')=(0,0)$  and the amplitude has the corresponding structure of a CFT three-point function. Incorporating a 2D shadow transform on one of the scalars scattering on the background, reveals an underlying CFT two-point function in the form of the shadow inner product advocated for in \cite{Crawley:2021ivb}.

We generalise our results to the case of magnetically charged scalars. One motivation for this is to understand the holographic dual of S-duality. Abelian gauge theory has a well-known duality symmetry which can be extended to an SL$(2, \mathbb{Z})$ duality group e.g. \cite{Cardy:1981qy, Cardy:1981fd, Shapere:1988zv, Verlinde:1995mz, Witten:1995gf}. To fully study the duality transformations it is necessary to include a non-vanishing $\theta$ angle so that in the case of the free theory the Lagrangian can be written as 
\begin{equation}
\mathcal{L}=-\frac{1}{4e^2} F^{\mu\nu}F_{\mu \nu}-\frac{\theta}{32 \pi^2}{*F}^{\mu\nu} F_{\mu \nu}~,
\end{equation}
where we have the field strength $F_{\mu \nu}=\partial_\mu A_\nu -\partial_\nu A_\mu$ and
the dual field strength 
$
{*F}_{\mu \nu}=\tfrac{1}{2}\epsilon_{\mu\nu\rho \sigma}F^{\rho \sigma}$~. Introducing a complex coupling $
\tau = \frac{\theta}{2\pi}+\frac{4\pi i}{e^2}
$
the Lagrangian can be written as
\begin{equation}
\mathcal{L}
=-\frac{i}{16\pi} \Big[\bar{\tau} F^{+\mu\nu}F^+_{\mu\nu} -\tau F^{-\mu\nu}F^-_{\mu\nu}\Big] ~
\end{equation}
where $F^\pm_{\mu\nu}=\tfrac{1}{2} F_{\mu\nu}\mp \tfrac{i}{2}{*F}_{\mu\nu}$ is the (anti) self-dual field strength combination.  The duality group is generated by
 shift transformations, which send $\theta \to \theta+2\pi n$, $n\in \mathbbm{Z}$
and the S-transformation which sends $\tau\to -1/\tau$. These transformations can be combined into the full SL$(2, \mathbbm{Z})$ group of transformations under which the coupling transforms as 
\begin{equation}
\tau\to \frac{a\tau+b}{c\tau+d}
\end{equation}
with $a, b, c, d \in \mathbbm{Z}$ and $ad-bc=1$. \footnote{In the Euclidean free theory, defined on a general four-manifold, the partition function is not invariant but rather transforms as a modular form
	\begin{equation}
	Z(\tau')=(c \tau +d)^{(\chi+\sigma)/4}(c \bar{\tau}+d)^{(\chi-\sigma)/4)} Z(\tau)
	\end{equation}
	where $\chi$ is the Euler characteristic and $\sigma$ the signature of the manifold \cite{Witten:1995gf}. } 
In the presence of charges, S-duality will interchange the electric and magnetic charges and
under a general SL$(2, \mathbbm{Z})$ transformation the magnetic charge $g_m$ and electric charge $g_e$ will transform as 
\begin{equation}
    g_m\to d g_m-c g_e~,~~~g_e \to a g_e -b   g_m~.
\end{equation}
The S-duality transformation can be viewed as mapping a configuration of the gauge potential into its magnetic dual. 
To compute the scattering amplitudes we need a formulation of Maxwell theory in the presence of magnetic charges. There are a number of formulations of scalar QED with both electric and magnetically charged particles \cite{Schwinger:1966nj, Zwanziger:1970hk, Schwarz:1993vs}. We will use the one-potential formulation \cite{Blagojevic:1983ft} for scalar QED, which is based on the earlier work for spinor QED \cite{Blagojevic:1978zv, Blagojevic:1979bm}. \footnote{This is equivalent to the two-potential formulation of Zwanziger \cite{Zwanziger:1970hk} which has been shown to be self-dual \cite{Csaki:2010rv}, \cite{Terning:2020dzg} under the SL$(2, \mathbb{Z})$ duality transformations.} In this formulation the magnetically charged scalar couples to the magnetic dual of the gauge potential and, to leading order, the scattering amplitude of a magnetically charged scalar in a background potential is the same as the amplitude for an electric scalar in the dual background. We use this to compute a range of two-point functions in non-trivial backgrounds for both electric and magnetic scalars.

 The magnetic dual of the spin-one CPW is a pseudo-conformal primary wavefunction (PCPW), which transforms as a pseudo-vector under Lorentz transformations. We compute the two-point function on this background and find that the amplitude is related to the electric amplitude by a T-duality like transformation. We find similar results for the conformally soft background. It is known that S-duality in four-dimensional gauge theories, dimensional reduced to two-dimensions, becomes T-duality \cite{Bershadsky:1995vm,Harvey:1995tg} and so it is natural to conjecture that in the two-dimensional celestial boundary dual that S-duality becomes T-duality. This has indeed been noted previously, \cite{Strominger:2015bla}, \cite{Kapec:2021eug}, and here we see it for scattering in non-trivial backgrounds. Two-dimensional T-duality can be understood as a change of variables in a two-dimensional sigma-model action \cite{Buscher:1987qj} and this can also be seen in the vacuum sector of the gauge theory.  For the Goldstone mode where the gauge potential is given in terms of a two-dimensional scalar field, the boundary 
 conditions can be written in terms of the two-dimensional Floreanini-Jackiw action \cite{Floreanini:1987as}. In section \ref{sec:bdact},
we show that the two-dimensional T-duality transformation corresponds to S-duality on the gauge field. A similar scalar action was introduced in \cite{Kapec:2021eug}, see also \cite{Kapec:2022hih, He:2024ddb}, to reproduce the known soft factors and IR divergences of QED. T-duality for this action similarly corresponds to an S-duality transformation in the effective coupling $\tau_{\text{eff}}$, which has logarithmic dependence on the IR and UV cut-offs, and an exchange of electric and magnetic currents. 

We further show that for magnetically charged shockwaves the celestial two-point function can again be recast as a three-point function but that the Dirac string position explicitly appears in the perturbative amplitude. This explicit dependence is not surprising as a similar phenomenon is known to occur in perturbative scattering between electrically and magnetically charged particles e.g. \cite{Rabl:1969gx, Terning:2018udc}. The perturbative approach is inconsistent with the Dirac quantisation condition and the issue can only be addressed after a re-summation of the perturbative expansion. Such a re-summation can be carried out for the IR divergent part of the amplitude \cite{Terning:2018udc} where it can be shown that the dependence on the Dirac string drops out. We find essentially the same result for the infrared sector of an electrically charged scalar wave scattering on a magnetically charged background. The infrared divergences exponentiate into a divergent phase given by a Wilson loop which can be shown to be independent of the Dirac string by the same argument as \cite{Terning:2018udc}. For a magnetically charged scalar in an electrically charged background we instead find the IR phase has the form of an 't Hooft loop which is consistent with previous studies of amplitudes with magnetically charged particles \cite{Choi:2019sjs}.

Certain backgrounds can be viewed as being generated by a point-particle source and so there is a interesting connection between $1\to1$ scattering from non-trivial backgrounds and $2\to 2$ particle scattering. In the gravitational case there is the well-known result of ’t Hooft \cite{tHooft:1987vrq} that the Aichelburg-Sexl shockwave solution can be used
to compute the scattering of high-energy, gravitationally interacting scalar particles. More generally, in \cite{Adamo:2021rfq} it was shown that for any stationary, linearised spacetime, the scalar 
two-point amplitude  has the form of an eikonal amplitude. In the context of celestial amplitudes \cite{deGioia:2022fcn} gave an explicit map between four-point amplitudes and scattering on specific shockwave backgrounds. In section \ref{sec:eik}, we generalise this result for the case of magnetically and electrically charged scalars showing that the background calculation can be related to the celestial four-point amplitude with both electric and magnetically charged scalars. 

This paper is organised as follows. In section \ref{sec:dual_sol} we construct magnetic dual conformal primary solutions and their conformally soft limits, magnetic and dyonic shockwaves and dynonic conformal primary wavefunctions. We then compute scalar two-point functions on these backgrounds in section \ref{sec:bgscat} and provide evidence that the holographic dual of bulk S-duality is boundary T-duality. In section \ref{sec:eik} we perform the all-order eikonal re-summation for magnetically charged scalars in terms of 't Hooft loops which are S-dual to the Wilson loops arising for electrically charged scalars. We also show that in the eikonal limit a concrete identification between four point amplitudes and two-point amplitudes in backgrounds can be extended to magnetically charged scalars. In section \ref{sec:bdact} we discuss the appearance of T-duality from S-duality in boundary actions. We conclude in section \ref{sec:outlook} with a list of open questions. Conventions and additional details are provided in Appendices \ref{app:Conventions}~--~\ref{integralcomp}.

%%%%%%%%%%%%%%%%%%%%%%%%%%%%%%%%%%%%%%%%%%%%%%%%%%%%%%%%%%%%%%%%%%%%%%%%%
%%%%%%%%%%%%%%%%%%%%%%%%%%%%%%%%%%%%%%%%%%%%%%%%%%%%%%%%%%%%%%%%%%%%%%%%%
\section{Dual solutions}
%%%%%%%%%%%%%%%%%%%%%%%%%%%%%%%%%%%%%%%%%%%%%%%%%%%%%%%%%%%%%%%%%%%%%%%%%
%%%%%%%%%%%%%%%%%%%%%%%%%%%%%%%%%%%%%%%%%%%%%%%%%%%%%%%%%%%%%%%%%%%%%%%%%
\label{sec:dual_sol}
In \cite{Witten:1995gf}, Witten gave an elegant path-integral derivation of S-duality in Abelian gauge theory. This was done using methods analogous to those used in deriving T-duality as a  world-sheet duality in two-dimensional sigma-models. Introducing an additional two-form field, $G_{\mu\nu}$, and a modified field strength, $\mathcal{F}_{\mu\nu}=F_{\mu \nu}-G_{\mu \nu}$, one can define an extended theory with the Lagrangian
\begin{align}
\label{eq:ext_lag}
\mathcal{L}= -\frac{i}{16\pi} \Big[-i\epsilon_{\mu \nu \rho \sigma} W^{\mu \nu} G^{\rho \sigma}+ \bar{\tau} \mathcal{F}^{+\mu\nu}\mathcal{F}^+_{\mu\nu} -\tau \mathcal{F}^{-\mu\nu}\mathcal{F}^-_{\mu\nu}\Big]
\end{align}
where $W_{\mu \nu}=\partial_{\mu}B_\nu-\partial_\nu B_\mu$ is the S-dual field strength in terms of the potential $B$. This theory is invariant under the extended gauge symmetry
\begin{equation}
\label{eq:ext_gi}
A \to A +\Xi~, ~~~G\to  G+d\Xi~,
\end{equation}
where $\Xi$ is a one-form gauge parameter, and under the dual gauge transformations $B \to B+d{\Lambda}$. In \cite{Witten:1995gf} a quantum argument is provided, see also \cite{Verlinde:1995mz,Lozano:1995aq}, that the extended theory \eqref{eq:ext_lag} is equivalent to both the original Maxwell theory and the S-dual version though for our purposes classical considerations are sufficient. By integrating out $G$, or equivalently by imposing the equations of motion
\begin{align}
\label{eq:dual_fs}
W^+_{\mu\nu}=\bar{\tau} \mathcal{F}^+_{\mu\nu}~, ~~~W^-_{\mu\nu}={\tau} \mathcal{F}^-_{\mu\nu}
\end{align}
and using the extended gauge invariance  \eqref{eq:ext_gi}
to set $A=0$ one finds
\begin{equation}
\mathcal{L}=-\frac{i}{16\pi} \Big[\left(-\frac{1}{\bar{\tau}}\right) W^{+\mu\nu}W^+_{\mu\nu} -\left(-\frac{1}{{\tau}}\right) W^{-\mu\nu}W^-_{\mu\nu}\Big]~
\end{equation}
which is the same as the original Maxwell Lagrangian with $A\to B$ and $\tau\to -1/\tau$. 

We can view the duality transformation in a different but related manner, as a map between gauge field configurations. In the case where $\theta=0$ the relation \eqref{eq:dual_fs}, for $G=0$, implies
\begin{equation}
\label{eq:duality_map}
W_{\mu\nu}=-\frac{4\pi}{e^2} {*F}_{\mu \nu}~.
\end{equation}
 Now, given a potential $A$ satisfying $F=dA$ we can attempt to find  an S-dual potential $B$ i.e.
\begin{align}
\label{eq:dual}
\partial_\mu B_\nu -\partial_\nu B_\mu=-\frac{4\pi }{e^2}\epsilon_{\mu \nu \rho \sigma} \partial^\rho A^\sigma.
\end{align}
Going to momentum space, introducing an arbitrary four-vector $n_\mu$ such that $n\cdot p\neq 0$, and choosing a gauge $n\cdot B=0$, we can ``solve" \eqref{eq:dual} for the magnetic dual potential,  see for example \cite{Terning:2020dzg},
\begin{align}\label{Bsolution}
\bar{B}{}_\mu(p)~``=" ~\frac{4\pi}{e^2}\frac{\epsilon_{\mu \nu \rho \sigma}n^\nu p^\rho \bar{A}^\sigma(p)}{n\cdot p}~
\end{align}
which depends on the choice of vector $n$.  Of course, as can be seen from direct substitution, this potential generically does not satisfy \eqref{eq:dual},  as we have projected the full set of equations onto the $n$ direction, but rather 
\begin{align}
\label{eq:dual_mod}
p_\mu \bar{B}{}_\nu-p_\nu \bar{B}{}_\mu=-\frac{4\pi}{e^2}\epsilon_{\mu\nu \rho \sigma} \left( p^\rho \bar{A}^\sigma-\frac{p^2}{n\cdot p} n^\rho \bar{A}^\sigma+\frac{p\cdot \bar{A}}{n\cdot p} n^\rho p^\sigma \right)~.
\end{align}
Only for potentials satisfying the vacuum wave equation $p^2 \bar{ A}^\sigma=0$ and the Lorenz condition $p_\mu \bar{{A}}^\mu=0$ does \eqref{eq:dual} hold. 
We will encounter such potentials below.

However, even if these conditions are not satisfied, the dual ``solution" \eqref{Bsolution} may be simply viewed as as a map between field configurations.
That is, given any potential $A_\mu$  we can define a magnetic dual potential 
\begin{align}
\label{eq:dual_pot}
\bar{A}^{{M}}{}_\mu(p)\equiv \frac{\epsilon_{\mu \nu \rho \sigma }n^\nu p^\rho \bar{{A}}^\sigma(p)}{n\cdot p}~
\end{align}
independently of any specific dynamics of the fields. 
If the potential ${A}^\mu$ in fact satisfies the vacuum Maxwell equation of motion, $\partial_\mu {F}^{\mu \nu}=0$, then it is easy to check that the dual potential $A^M_\mu$ satisfies the dual equation of motion $\partial^\mu {F}^M_{\mu \nu}=0$. Furthermore, as the dual potential automatically satisfies the Lorenz condition $\partial^\mu A^M_\mu=0$, it consequently satisfies the vacuum wave equation $\Box A^M_\mu=0$. 

%%%%%%%%%%%%%%%%%%%%%%%%%%%%%%%%%%%%%%%%%%%%%%%%%%%%%%%%%%%%%%%%%%%%%%%%%
\subsection{Magnetic conformal primary waves}
%%%%%%%%%%%%%%%%%%%%%%%%%%%%%%%%%%%%%%%%%%%%%%%%%%%%%%%%%%%%%%%%%%%%%%%%%

While the dual potential does not solve \eqref{eq:dual} unless the potential satisfies the source free Maxwell equations in Lorenz gauge,
both of these conditions are naturally satisfied by the photon conformal primaries defined in \cite{Pasterski:2017kqt}. As a result,  their magnetic duals which we construct in the following will also satisfy the vacuum Maxwell equations. It will be convenient to define a map between $\mathbb R^2$ and $\mathbb{R}^{1,3}$ given by the null vector 
\begin{align}
\label{eq:qmom1}
q^\mu=(1+|\vec{z}|^2, 2 \vec{z}, 1-|\vec{z}|^2 ),
\end{align}
in terms of the two-dimensional vector $\vec{z}$ as well as its projection
\begin{align}
\partial_a q^\mu\equiv \frac{\partial q^\mu}{\partial z^a}=2(z^a, \delta^{a1}, \delta^{a2},-z^a),
\end{align}
where $a=1,2$ is an index for the two-dimensional space for which we can also introduce a complex coordinate $z=(z^1+i z^2)$. While $\vec z$ transforms non-linearly under an SO(1,3) transformation, its embedding $q$ transforms linearly,
\begin{align}
q^\mu(\vec{z}\,{}')=\left|\frac{\partial \vec{z}\,'}{\partial \vec{z}}\right|^{1/2} \Lambda^\mu{}_\nu q^\nu(\vec{z}),
\end{align}
where $\Lambda^\mu{}_\nu$ is the usual vector representation (see e.g. \cite{Pasterski:2017kqt}).
We will briefly review the electric solutions constructed in \cite{Pasterski:2017kqt} as vectors in $\mathbb{R}^{1,3}$ and spin-one conformal primaries in $\mathbb{R}^2$ before constructing their magnetic dual solutions.

%%%%%%%%%%%%%%%%%%%%%%%%%%%%%%%%%%%%%%%%%%%%%%%%%%%%%%%%%%%%%%%%%%%%%%%%%
\paragraph{Electric primary}
%%%%%%%%%%%%%%%%%%%%%%%%%%%%%%%%%%%%%%%%%%%%%%%%%%%%%%%%%%%%%%%%%%%%%%%%%

The photon conformal primary gauge potential, which is labelled by a two-dimesnional polarisation index $a$ in addition to a four-dimensional vector index,  is given by \cite{Pasterski:2017kqt}
\begin{align}
\label{eq:cpw}
A^{\Delta, \pm}_{\mu; a}=V^{\Delta, \pm}_{\mu;a}+\partial_\mu \alpha^{\Delta, \pm}_{a},
\end{align}
with  
\begin{align}\label{Valpha}
V^{\Delta, \pm}_{\mu; a}=\frac{\Delta-1}{\Delta}\frac{\partial_a q_\mu}{(-q\cdot x_\pm)^{\Delta}}~,~~~\text{and}~~~ \alpha^{\Delta, \pm}_{a}=\frac{1}{\Delta}\frac{\partial_a q\cdot x_\pm}{(-q\cdot x_\pm)^{\Delta}}~,
\end{align}
where we regulate the zero of the denominator $q\cdot x_\pm$  by defining $x^\mu_\pm=x^\mu\pm i \epsilon \delta^\mu_0$.
The gauge potentials $A_{\mu;a}^{\Delta,\pm}$ are solutions to the source-free Maxwell equations and satisfy both the Lorenz and radial gauge conditions 
\begin{align}
x^\mu A_{\mu;a}^{\Delta,\pm}=\partial^\mu A_{\mu;a}^{\Delta,\pm}=0~.
\end{align}
They were found in \cite{Pasterski:2017kqt} from the spin-one bulk-to-boundary propagator in $H_{3}$ whose uplift to $\mathbbm{R}^{1,3}$ is given by
\begin{align}
\label{eq:G}
G^\Delta_{\mu;\nu} =\frac{(-q\cdot \hat p)\eta_{\mu\nu}+q_\mu \hat p_\nu}{(-q\cdot \hat{p})^{\Delta+1}}
\end{align}
where $\hat{p}$ is a time-like vector and $q$ the same null vector as before. The uplifted propagator is transverse in the sense that
\begin{align}
\hat{p}^\mu G^\Delta_{\mu;\nu}=q^\nu G^{\Delta}_{\mu;\nu}=0~.
\end{align}
Under an SO(1,3) transformation $G^\Delta_{\mu;\nu}(\hat{p};q)$ transforms as a rank-two tensor in four dimensions and as a scalar conformal primary in two dimensions:
\begin{align}
G_{\mu;\nu}^\Delta(\Lambda \hat{p}; \Lambda q)=\left|\frac{\partial \vec{z}\,'}{\partial \vec{z}}\right|^{-\Delta/2} \Lambda_\mu{}^\rho \Lambda_\nu{}^\sigma G^\Delta_{\rho ;\sigma}(\hat{p}; q)~.
\end{align}
Its projection on the second index, $\frac{\partial q^\nu}{\partial z^a} G^\Delta_{\mu;\nu}$,  after the replacement $\hat p \to x$ , becomes the photon conformal primary \eqref{eq:cpw} which transforms in the desired way,
\begin{align}
A_{\mu; a}^\Delta(\Lambda x; \vec{z}\,')=\frac{\partial z^b}{\partial {z'}^a}\left|\frac{\partial \vec{z}\,'}{\partial \vec{z}}\right|^{-(\Delta-1)/2} \Lambda_\mu{}^\rho  A^\Delta_{\rho; b}(x; \vec{z})~,
\end{align}
that is as a vector in $\mathbb{R}^{1,3}$ and a spin-one conformal primary in $\mathbb{R}^2$.

Another important photon conformal primary solution is obtained from the shadow transform 
\begin{equation}
\label{eq:Shcpw}
     \widetilde{A^{\Delta,\pm}_{\mu;\nu}}(x;q)=\frac{-\Delta}{\pi}\int d^2\vec{z}\,'\frac{(-\frac{1}{2}q\cdot q')\delta^\rho_\nu+\frac{1}{2}q_\nu'q^\rho}{(-\frac{1}{2}q\cdot q')^{3-\Delta}}A^{\Delta,\pm}_{\mu;\nu}(x;q')
\end{equation}
of the uplifted primary $A^{\Delta,\pm}_{\mu;\nu}(x;q)=G^{\Delta,\pm}_{\mu;\nu}(x;q)$ \cite{Pasterski:2017kqt}. The normalisation of the shadow for spin-one primaries
is chosen such that the two-dimensional shadow transform squares to 1 (see Appendix D of \cite{Pasterski:2021fjn}). 
After projection we obtain the shadow conformal primary \footnote{For complex coordinates in the two-dimensional space, i.e. $a=z,\bar z$, the primary on the rhs of \eqref{eq:Shcpw} has an extra complex conjugation on this label since the two-dimensional spin gets flipped under the shadow transform -- see e.g. Appendix~D of \cite{Pasterski:2021fjn}.}
\begin{equation}
\widetilde{A^{\Delta,\pm}_{\mu;a}}=-(-x_\pm^2)^{1-\Delta} A^{2-\Delta,\pm}_{\mu;a}
\end{equation}
which also solves the source-free Maxwell equations and satisfies Lorenz and radial gauge.

%%%%%%%%%%%%%%%%%%%%%%%%%%%%%%%%%%%%%%%%%
\paragraph{Magnetic primary}
%%%%%%%%%%%%%%%%%%%%%%%%%%%%%%%%%%%%%%%%%
We now define a magnetic dual of the photon conformal primaries as
\begin{align}
\label{eq:AbarM}
\bar{A}^{M, \Delta, \pm}_{\mu; a}=\frac{\epsilon_{\mu \nu \rho \sigma} n^\nu p^\rho \bar{A}^{\Delta, \pm}_{a}{}^\sigma}{n\cdot p}=\frac{\epsilon_{\lambda \kappa \mu \nu} n^\nu p^\rho \bar{V}^{\Delta, \pm}_{a}{}^\sigma}{n\cdot p}
\end{align}
which again depend on the position of the Dirac string given by the vector $n$. 
To find the explicit expression, first notice that the potential in \eqref{Valpha} is the Mellin transform of a plane wave
\begin{equation}
    V^{\Delta,\pm}_{\mu;a}=\frac{(\pm i)^\Delta}{\Gamma(\Delta)} \frac{\Delta-1}{\Delta} \partial_a q_\mu \int_0^\infty d\omega \omega^{\Delta-1} e^{\pm i\omega q\cdot x-\epsilon \omega},
\end{equation}
and the Fourier transform of the latter yields a Dirac-delta.
Going back to position space and the conformal primary basis we find
\begin{align}
{A}^{M, \Delta, \pm }_{\mu; a}&=(2\pi)^4 \frac{(\pm i)^\Delta}{\Gamma(\Delta)}  \frac{\Delta-1}{\Delta}\frac{\epsilon_{\mu \nu \rho \sigma} n^\nu q^\rho \partial_a q^\sigma }{n\cdot q}\int \frac{d^4p}{(2\pi)^4}
\int_0^\infty d\omega
~ \omega^{\Delta-1}\delta^{(4)}(p\mp \omega q)e^{ip\cdot x-\epsilon \omega}\nn\\
&= \frac{\Delta-1}{\Delta}\frac{\epsilon_{\mu \nu \rho \sigma} n^\nu q^\rho \partial_a q^\sigma }{n\cdot q}\frac{1}{(-q\cdot x_\pm )^\Delta}~.
\end{align}
The magnetic dual potential can thus be expressed as
\begin{equation}
\label{eq:dual_mom_cpw}
{A}^{M, \Delta, \pm }_{\mu; a}=\frac{\epsilon_{\mu \nu \rho \sigma} n^\nu q^\rho {A}^{\Delta, \pm}_{a}{}^\sigma}{n\cdot q}=\frac{\epsilon_{\mu \nu \rho \sigma} n^\nu q^\rho {V}^{\Delta, \pm}_{a}{}^\sigma}{n\cdot q}.
\end{equation} 
These potentials are vacuum solutions of the dual Maxwell equations satisfying 
\begin{align}
\partial^\mu F^{M, \Delta \pm}_{\mu\nu;a}=\Box{A}^{M, \Delta, \pm}_{\nu; a}=0~
\end{align} 
and additionally they do in fact satisfy
\begin{align}
\epsilon_{\mu\nu}{}^{\rho \sigma} \partial_\rho A^{M, \Delta \pm}_{ \sigma;a}
&=\partial_\mu A^{\Delta, \pm}_{\nu; a}-\partial_\nu A^{\Delta, \pm}_{\mu; a}\nn\\
&
=(q_\mu \partial_a q_\nu-q_\nu \partial_a q_\mu)\frac{\Delta-1}{(-q\cdot x_\pm)^{\Delta+1}}
~.
\end{align}

As a result, while the dual potentials have explicit dependence on the vector $n$, it is clear that the dual field strengths do not,
\begin{align}
F^{M, \Delta \pm}_{\mu\nu;a}=(1-\Delta) \frac{\epsilon_{\mu\nu\rho\sigma} q^\rho \partial_a q^\sigma}{(-q\cdot x\mp i \epsilon)^{\Delta+1}}~,
\end{align}
and so we can find a gauge equivalent potential which doesn't depend on $n$ such as
\begin{align}
\label{eq:ge_dpot}
A_{\mu; a}^{M, \Delta, \pm}=\frac{\epsilon_{\mu\nu\rho \sigma} \partial_a q^\nu x^\rho q^\sigma}{(-q\cdot x_\pm)^{\Delta+1}}~.
\end{align}
This solution again satisfies Lorenz and radial gauge.  Comparing this to
\begin{equation}
    A^\Delta_{\mu;a}=\frac{q_\mu \partial_a q\cdot x-\partial_a q_\mu q\cdot x}{(-q\cdot x)^{\Delta+1}},
\end{equation}
we see that the effect of dualisation is the replacement  
\begin{equation}
    q_\mu \partial_a q\cdot x-\partial_a q_\mu q\cdot x\to \epsilon_{\mu\nu\rho \sigma} \partial_a q^\nu x^\rho q^\sigma.
\end{equation}
The uplift of this dual potential,
\begin{align}
\label{eq:Gpseudo}
G^{M,\Delta}_{\mu;\nu} =\frac{\epsilon_{\mu \nu \rho \sigma} \hat{p}^\rho q^\sigma}{(-q\cdot \hat{p})^{\Delta+1}},
\end{align}
has a similar interpretation as its electric counterpart \eqref{eq:G}, but now as a rank-two pseudo tensor in $\mathbbm{R}^{1,3}$ and a scalar conformal primary in $\mathbbm{R}^{2}$ as can be seen from its SO(1,3) transformation
\begin{align}
G_{\mu;\nu}^{M,\Delta}(\Lambda \hat{p}; \Lambda q)=\left|\frac{\partial \vec{z}\,'}{\partial \vec{z}}\right|^{-\Delta/d} \text{det}(\Lambda) \Lambda_\mu{}^\rho \Lambda_\nu{}^\sigma G^{M,\Delta}_{\rho; \sigma}(\hat{p}; q)~.
\end{align}
We can convert this to a conformal primary wavefunction by the projection 
\begin{align}
\frac{\partial q^\nu}{\partial z^a} G^{M,\Delta}_{\mu;\nu}=\frac{\epsilon_{\mu\nu\rho\sigma} \partial_a q^\nu \hat{p}^\rho q^\sigma}{(-q\cdot \hat{p})^{\Delta+1}}~,
\end{align}
which can be seen to correspond to the potential \eqref{eq:ge_dpot} under the replacement $\hat{p}\to x$.
The transformation properties of this magnetic solution 
as a pseudo-vector in $\mathbb{R}^{1,3}$ and a spin-one conformal primary in $\mathbb{R}^2$,
\begin{equation}
A_{\mu; a}^{M,\Delta}(\Lambda x; \vec{z}\,')=\frac{\partial z^b}{\partial {z'}^a}\left|\frac{\partial \vec{z}\,'}{\partial \vec{z}}\right|^{-(\Delta-1)/2}\text{det}(\Lambda) \Lambda_\mu{}^\rho  A^\Delta_{\rho; b}(x; \vec{z})~,
\end{equation}
are guaranteed by those of $G^{M,\Delta}_{\mu;\nu}$, $\partial_a q^\nu$ and the transversality
\begin{equation}
    \hat{p}^\mu G^{M,\Delta}_{\mu;\nu}=q^\nu G^{M,\Delta}_{\mu;\nu}=0.
\end{equation}
As a result we see that dualisation maps the conformal primary photon to a dual pseudo-conformal primary photon. As for the former we can define a new magnetic solution from the shadow transform of the magnetic primary \eqref{eq:ge_dpot}. The result is \footnote{Again, for $a=z,\bar z$, the primary on the rhs of \eqref{eq:MShcpw} has an extra complex conjugation on this label.}
\begin{equation}
\label{eq:MShcpw}
\widetilde{A^{M,\Delta,\pm}_{\mu;a}}=- (-x_\pm^2)^{1-\Delta} A^{M,2-\Delta,\pm}_{\mu;a},
\end{equation}
which also solves the source-free Maxwell equations and satisfy Lorenz and radial gauge.

%%%%%%%%%%%%%%%%%%%%%%%%%%%%%%%%%%%%%%%%%%%%%%%%%%%%%%%%%%%%%%%%%%%%%%%%%
\subsection{Magnetic logarithmic mode}
%%%%%%%%%%%%%%%%%%%%%%%%%%%%%%%%%%%%%%%%%%%%%%%%%%%%%%%%%%%%%%%%%%%%%%%%%
Another interesting photon primary is obtained by combining primaries and shadows in a limit where they have the same conformal dimension. For $\Delta=1$ this reveals a new logarithmic branch of the solution space \cite{Donnay:2018neh}.
  
\paragraph{Electric logarithmic mode}
The $\Delta=1$ logarithmic mode  of \cite{Donnay:2018neh} is defined as
\begin{align}
\label{eq:log}
{A}^{\text{log}, \pm}_{\mu; a}
&\equiv \lim_{\Delta\to 1} \partial_\Delta(A_{\mu; a}^{\Delta, \pm}+\widetilde{A^{2-\Delta, \pm}_{\mu; a^*}})\nn\\
&=-\text{log}[-x_\pm^2]A^{1,\pm}_{\mu;a}~.
\end{align}
For the purpose of constructing the magnetic dual potential, it will be convenient to write this as
 \begin{align}
  {A}^{\text{log}}_{\mu; a}   &={V}^{\text{log}}_{\mu; a}+\partial_\mu \alpha^{{\text{log}}}_{ a},
 \end{align}
where 
 \begin{equation}
 \label{Vlog}
     {V}^{\text{log}}_{\mu; a}=-\frac{2x_\mu}{x^2}\partial_a \text{log}[-q\cdot x],
    \end{equation}
and we can henceforth ignore the pure gauge term $\partial_\mu \alpha_a^{\log}$. Here we have also dropped the regularisation $x_\pm$.
Finding the momentum space representation depends strongly on the singularities of the $1/x^2$ term (see Appendix \ref{app:log}) where using $x_\pm$ corresponds to using retarded/advanced prescription. However, it is interesting to consider different prescriptions, in particular the Feynman prescription which is natural from an amplitudes perspective. 

\paragraph{Magnetic logarithmic mode}
Using the Feynman prescription for the singularities we have
\begin{align}
\bar{{V}}^{\text{log},F}_{\mu; a}(p)=2(2\pi)^2 
\frac{p\cdot q \partial_a q_\mu - q_\mu \partial_a q\cdot p}{p^2 (p\cdot q)^2},
\end{align}
up to terms proportional to $p_\mu$,
so that 
\begin{align}
\label{eq:AMlogF}
\bar{{A}}^{M,\text{log},F}_{\lambda; a}(p)&=2(2\pi)^2 
\epsilon_{\lambda \kappa \mu \nu} n^\kappa p^\mu \frac{p\cdot q \partial_a q^\nu - q^\nu \partial_a q\cdot p}{ p^2(n\cdot p) (p\cdot q)^2}\nn\\
&=2(2\pi)^2
\epsilon_{\lambda \kappa \alpha \beta} \left(\frac{n^\kappa}{n\cdot p}- \frac{p^\kappa}{p^2}\right)  \frac{ q^\alpha \partial_a q^\beta}{  (p\cdot q)^2}~.
\end{align}
Here we see that the magnetic dual logarithmic potential depends explicitly on the location of the Dirac string. However, we may make the gauge choice $n^\mu = q^\mu$ such that
\begin{align}
\label{eq:AMlogFgf}
\bar{{A}}^{M,\text{log},F}_{\mu; a}(p)
&=2(2\pi)^2
 \frac{\epsilon_{\mu \nu \rho \sigma} p^\nu \partial_a q^\rho q^\sigma}{p^2  (p\cdot q)^2},
\end{align}
where we see that the effect of  dualisation is a simple replacement of the numerator factor. 

Rather than dualising the electric logarithmic mode, we could apply a similar analysis as in \cite{Donnay:2018neh} directly to the magnetic primary \eqref{eq:ge_dpot} and define the magnetic logarithmic mode as
\begin{align}
\label{eq:AMlognon}
    {A}^{M,\rm{log}, \pm}_{\mu; a}&\equiv\lim_{\Delta\to 1} \partial_\Delta \left({A}^{M,\Delta, \pm}_{\mu; a}-\widetilde{A^{M,2-\Delta, \pm}_{\mu; a^*}}\right)\nn\\
    &=-\log(-x^2_\pm) {A}^{M,1, \pm}_{\mu; a}.
\end{align}
This expression has the benefit of being manifestly independent of the Dirac string location.

%%%%%%%%%%%%%%%%%%%%%%%%%%%%%%%%%%%%%%%%%%%%%%%%%%%%%%%%%%%
\subsection{Magnetic conformally soft mode}
\label{CSmodes}
%%%%%%%%%%%%%%%%%%%%%%%%%%%%%%%%%%%%%%%%%%%%%%%%%%%%%%%%%%%
A special case of the conformal primary \eqref{eq:cpw} and its magnetic dual \eqref{eq:ge_dpot} arises for $\Delta=1$.
\paragraph{Electric memory mode}
The photon conformal primary \eqref{eq:cpw} with $\Delta=1$ is pure gauge. The canonical partner \footnote{with respect to the inner product $i(A,A')=\int d^3x\left(A^\mu {F'}^*_{0\mu}-{A'}^{*\mu}F_{0\mu} \right)$} of this Goldstone mode is  the $\Delta=1$ Memory mode. To identify this mode a new conformally soft primary given by 
\begin{equation}
\label{eq:ACS}
A^{CS}_{\mu;a} =A^{1}_{\mu;a}\left[\Theta(x^2)+\log(x^2)(q\cdot x) \delta(q\cdot x)\right]
\end{equation}
was constructed in \cite{Donnay:2018neh} from a combination of incoming and outgoing logarithmic modes.
 As in \cite{Pasterski:2021dqe} we may split \eqref{eq:ACS} %into a shift and shockwave type contribution 
into the two contributions
\begin{equation}
\label{CSsplit}
    A^{CS'}_{\mu;a}(x;q) =\Theta(x^2)A^{1}_{\mu;a}
   % =\Theta(x^2)\partial_a\left[\frac{q_\mu}{-q\cdot x}\right]
   ,\quad 
    A^{CS''}_{\mu;a}(x;q) =\log(x^2) (q\cdot x)\delta(q\cdot x) A^{1}_{\mu;a}.
\end{equation}
This split highlights that it is the $CS'$ mode that is responsible for the pairing
\begin{equation}
    i(A^1_a,A^{CS'}_{\bar a'})=-8\pi^2 \delta^{(2)}(z-z'),
\end{equation}
as was also observed in \cite{Arkani-Hamed:2020gyp}. Still the two contributions share similar features as wavefunctions and as backgrounds for amplitudes.
After a bit of massaging the split \eqref{CSsplit} furthermore reveals that we may write these primaries as descendants
\begin{equation}
\label{shiftshockdescendants}
A^{CS'}_{\mu;a}=\partial_a A^{shift}_\mu, \quad  A^{CS''}_{\mu;a}=\partial_a A^{shock}_\mu,
\end{equation}
of, respectively, a shift and shockwave potential
\begin{equation}
\label{shiftshock}
    A^{shift}_\mu=-q_\mu\Theta(x^2) (q\cdot x)^{-1}, \quad A^{shock}_\mu=-q_\mu \log(x^2) \delta(q\cdot x),
\end{equation}
where we used $\delta'(x)=-\delta(x)/x$ to obtain the latter. 
We will return to this below. 

The Fourier transform for the shift and shockwave potentials are
\begin{equation}
\label{shiftshockFourier}
    \bar A^{shift}_\mu=-8\pi^2 q_\mu \frac{\delta(p^2)}{q\cdot p}, \quad \bar A^{shock}_\mu=8\pi^2 q_\mu \frac{\delta(q\cdot p)}{p^2}.
\end{equation}
Their momentum-space descendants are given by
\begin{equation}
\label{eq:ACS'}
\bar{A}^{CS'}_{\mu;a}(p;q)=8\pi^2 (q_\mu \partial_aq\cdot p-\partial_a q_\mu q\cdot p)\frac{\delta(p^2)}{(q\cdot p)^2},
\end{equation}
and
\begin{equation}
\label{eq:ACS''}
\bar{A}^{CS''}_{\mu;a}(p;q)=8\pi^2 (q_\mu \partial_aq\cdot p-\partial_a q_\mu q\cdot p)\frac{\delta'(p\cdot q)}{p^2}~.
\end{equation}
The shift descendant potential $\bar{A}^{CS'}_{\mu;a}$ satisfies the Lorenz gauge condition and the equation of motion both in position-space as well as in momentum-space,
\begin{align}
p^\mu \bar{A}^{CS'}_{\mu;a}=0~,~~~
p^2 \bar{A}^{CS'}_{\mu;a}=0.
\end{align}
The situation is more subtle for the shockwave descendant potential $\bar{A}^{CS''}_{\mu;a}$. While it also satisfies the Lorenz gauge condition in momentum space its equation of motion is sourced
\begin{align}
\label{eq:shck_src}
p^\mu \bar{A}^{CS''}_{\mu;a}=0~,~~~p^2 \bar{A}^{CS''}_{\mu;a}=  \partial_a  (8\pi^2 q_\mu \delta(p\cdot q)).
\end{align}
This is of particular relevance in the current context as these source terms effect the nature of the dual potentials. The corresponding equation of motion in position space is 
\begin{align}
    \partial^\nu F^{CS''}_{\nu\mu;a}=\Box{A}^{CS''}_{\mu;a}=\partial_a\big[-4\pi  \int d\tau ~ q_\mu \delta^{(4)}(x-\tau q)\big]
\end{align}
where the source is the descendant of the well-known shockwave current which vanishes everywhere except along the world-line of a massless particle.

\paragraph{Magnetic memory mode}
As the shift potential $\bar{A}^{CS'}_{\mu;a}$ satisfies the Lorentz condition and equation of motion, the dualised potential will produce a field strength independent of the reference vector $n$ and so any dependence on $n$ is simply a gauge artefact. This is immediately apparent as
\begin{align}
\label{eq:mag_csp}
\bar{A}^{M,CS'}_{\mu;a}(p;q)&=8 \pi^2 \frac{\delta(p^2)}{(q\cdot p)^2(n\cdot p) } \epsilon_{\mu\nu\rho \sigma}\partial_a q^\rho q^\sigma [n^\nu p^2 -(n\cdot p) p^\nu] \nn\\
&=-8 \pi^2 \frac{\delta(p^2)}{(q\cdot p)^2 } \epsilon_{\mu\nu\rho \sigma}p^\nu \partial_a q^\rho q^\sigma,
\end{align}
where we have dropped terms proportional to $p_\mu$ and due to the Dirac-$\delta$ the dependence on $n$ drops out. Here we can see again that 
 the effect of dualisation for the potential is simply to make the substitution
\begin{align}
\label{eq:sub}
\partial_a q_\mu q\cdot p -q_\mu \partial_aq\cdot p\to \epsilon_{\mu\nu\rho \sigma} p^\nu \partial_a q^\rho q^\sigma~,
\end{align}
though in this case we didn't need to make a particular gauge choice. 
On the other hand, the magnetic dual for $CS''$ is given by 
\begin{align}
\label{eq:mag_cspp}
\bar{A}^{M,CS''}_{\mu;a}(p;q)&=8 \pi^2 \frac{\delta'(p\cdot q)}{ p^2(n\cdot p) } \epsilon_{\mu\nu\rho \sigma} \partial_a q^\rho q^\sigma[n^\nu p^2 -(n\cdot p) p^\nu] ~.
\end{align}
As can be seen by computing $\bar{F}^{M,CS''}_{\mu\nu;a}(p;q)$, the $n$ dependence is not a gauge artefact but arises from the non-trivial equations of motion. Computing the momentum space equation of motion, one also finds that this potential is no longer of vacuum form. However, if we again make the particular gauge choice $n^\mu=q^\mu$, where the string lies along the source trajectory, we find that the dual potential is again given by the substitution  \eqref{eq:sub}. 

Alternatively, starting from the magnetic primary \eqref{eq:ge_dpot} we can define a conformally soft magnetic primary that is manifestly independent of the Dirac string location as follows. First note that the field strength of the $\Delta=1$ magnetic primary vanishes and so the latter is a magnetic Goldstone mode. Its canonical partner can be constructed along similar lines as above. That is, following the definition of the electric conformally soft primary in \cite{Donnay:2018neh} we define the field strength of the magnetic conformal primary through a combination of incoming and outgoing modes
\begin{equation}
    F^{M, CS}_{\mu\nu;a}\equiv\frac{1}{2\pi i} \left( F^{M,\log,+}_{\mu\nu;a}-F^{M,\log,-}_{\mu\nu;a}\right),
\end{equation}
and subsequently take $\epsilon\to0$. 
The magnetic conformally soft primary is thus given by
\begin{equation}
\label{eq:MCS}
A^{M, CS}_{\mu;a} \equiv A^{M,1}_{\mu;a}\left[\Theta(x^2)+\log(x^2)(q\cdot x) \delta(q\cdot x)\right],
\end{equation}
with the $\Theta(x^2)$ term corresponding to the shift mode $CS'$ and the $\delta(q\cdot x)$ term yielding the shockwave type contribution $CS''$. One can check that the inner product of the $\Delta=1$ magnetic Memory mode \eqref{eq:MCS} and the $\Delta=1$ magnetic Goldstone mode takes the form
\begin{equation}
   % i(A^{M,1}_{a=z},A^{MCS}_{a'=\bar z'})=- i(A^{1}_{a=z},A^{CS}_{a'=\bar z'})=8\pi^2\delta^{(2)}(z-z'),
    i(A^{M,1}_{a},A^{M,CS'}_{\bar a'})=- i(A^{1}_{a},A^{CS'}_{\bar a'})=8\pi^2\delta^{(2)}(z-z'),
\end{equation}
thus confirming that they are canonically paired conformally soft modes.

%%%%%%%%%%%%%%%%%%%%%%%%%%%%%%%%%%%%%%%%%%%%%%%%%%%%%%%%%%%
\subsection{Magnetic shockwave}
%%%%%%%%%%%%%%%%%%%%%%%%%%%%%%%%%%%%%%%%%%%%%%%%%%%%%%%%%%%

The simplest example of a duality transformation is perhaps the case of the Coulomb potential. In that case the dual potential is that of the Dirac monopole, though with the Dirac string split into two. To find the magnetic shockwave we can consider performing an ultraboost on the monopole potential, see for example \cite{Argurio:2008nb, Bahjat-Abbas:2020cyb}, or more directly we can dualise the electromagnetic shockwave background. 
In momentum space the electric shockwave potential is given by
\begin{align}
\bar{{A}}_\mu^{E-shock}(p)=8\pi^2 q_\mu \frac{\delta(q\cdot p) }{p^2} 
\end{align}
where $q$ is a null vector and so
the dualised background is
\begin{align}
\label{eq:dual_emshock}
\bar{A}_\mu^{M-shock}(p) = 8\pi^2 \epsilon_{\mu \nu \rho \sigma} \frac{n^\nu p^\rho q^\sigma}{n\cdot p}\frac{\delta(p\cdot q)}{ p^2}~.
\end{align}
Let us take for example, $q_\mu=(-1,0,0,1)$ and $n_\mu=(0,1,0,0)$ so that the Dirac string is orthogonal to the shockwave direction. In this case
\begin{align}
\bar{A}_\mu^{M-shock}(p)=-8\pi^2 q_\mu \frac{p_2}{p_1}\frac{\delta(q\cdot p)}{p^2}
\end{align}
so that the position space potential is
\begin{align}\label{eq:mag_shock}
A_\mu^{M-shock}(x) = 2 q_\mu \tan^{-1}(x/y) \delta(t-z) ~.
\end{align}
This is essentially the single copy of the NUT-wave solutions first constructed in \cite{Argurio:2008nb} which is a boosted version of the Taub-NUT solution whose single copy is known to be the monopole potential \cite{Luna:2015paa}.

%%%%%%%%%%%%%%%%%%%%%%%%%%%%%%%%%%%%%%%%%%%%%%%%%%%%%%%%%%%
\subsection{Dyonic shockwaves from scattering amplitudes}
%%%%%%%%%%%%%%%%%%%%%%%%%%%%%%%%%%%%%%%%%%%%%%%%%%%%%%%%%%%

Recently, it was shown that shockwaves in gauge theories can be constructed from three-particle scattering amplitudes with one leg off shell \cite{Cristofoli:2020hnk}. It has also been shown that the deformation of three-particle amplitudes by a phase is a duality rotation \cite{Huang:2019cja,Moynihan:2020gxj,Emond:2020lwi,Emond:2021lfy}, converting charged particles into dyons. In this section, we will show that dyonic shockwaves can be constructed by duality-rotating a scalar QED three-particle amplitude with an off-shell leg. We begin by introducing an off-shell gauge field operator
\[
\mathbb{A}^\mu(x) = \sum_h\int \hat{d}^{4} p \frac{1}{p^{2}}\left(\varepsilon_h^\mu(p) a^h_pe^{-ip\cdot x} + \varepsilon_{-h}^\mu(p) a^{h\dagger}_p e^{ip\cdot x}\right),
\]
where we adopt the notation 
\[
\hat{d}^4p = \frac{d^4p}{(2\pi)^4}.
\]
We can evolve this with the $S$-matrix, evaluating its expectation value taken with respect to an initial state describing an on-shell massless scalar, given by 
\[
\ket{\Psi} = \int \hat{d}^4q\hat{\delta}(q^2) \varphi(q)\ket{q},
\]
where $\hat{\delta}(X) = \hat{\delta}^{(1)}(X)$, and we are using the general notation $\hat{\delta}^{(D)}(X) = (2\pi)^D\delta(X)$.

The $\varphi(q)$ above is a momentum-space wavefunction ensuring that the scalar is a classically localised source, whose precise details are unimportant -- for more details see ref. \cite{Kosower:2018adc}. Taking the expectation value with respect to these states then yields the potential expressed in terms of a three-particle amplitude in scalar QED, where the photon is off-shell
\[
\braket{\mathbb{A}^\mu(x)} &= \sum_h\int \hat{d}^{4} p \frac{1}{p^{2}}\left(\varepsilon_h^\mu(p) \braket{\Psi|S^\dagger a^h_p S|\Psi}e^{-ip\cdot x} + \varepsilon_{-h}^\mu(p) \braket{\Psi|S^\dagger a^{h\dagger}_p S|\Psi} e^{ip\cdot x}\right)\\
&= \sum_h\int \hat{d}^{4} p \frac{1}{p^{2}} \hat{d}^{4}q\hat{d}^{4}q'\hat{\delta}(q^2)\hat{\delta}(q'^2)|\varphi(q)|^2\\
&\kern+100pt \times\left(i\varepsilon_h^\mu(p) \braket{q|a^h_p T|q'}e^{-ip\cdot x} - i\varepsilon_{-h}^\mu(p) \braket{q|T^\dagger a^{h\dagger}_p|q'} e^{ip\cdot x}\right),
\]
where we have expanded $S = 1+iT$ and used $\varphi(q)\varphi^*(q') = |\varphi(q)|^2 + \cl{O}(\hbar)$. 
The three-particle amplitude --- with one leg off-shell --- is defined as
\[
\braket{q|a_p^{h}T|q'} = \braket{q,p^h|T|q'} = \cl{M}_3^\mu[q,q']\varepsilon_\mu^{-h}(p)\hat{\delta}^{(4)}(q+q'+p),
\]
and so we can expand the gauge field as
\[
\braket{\mathbb{A}^\mu(x)} &= \sum_h\int \hat{d}^{4} p \frac{1}{p^{2}} \hat{d}^{4}q\hat{d}^{4}q'\hat{\delta}(q^2)\hat{\delta}(q'^2)\hat{\delta}^{(4)}(q+q'+p)\\
& \kern+100pt \times \Im\left[|\varphi|^2\varepsilon^{h\mu}(p)\cl{M}_3[q,q',p^{-h}]e^{-ip\cdot x}\right].
\]
The amplitude for a scalar coupled to a photon of helicity $h$ and charge $Q$ is given by
\[
\cl{M}_3[q,q',p^h] = -iQ(q+q')\cdot\varepsilon^h(p) = \cl{M}_3^\mu[q,q']\varepsilon_\mu^h(p).
\]
At the level of three-particle amplitudes, duality transformations are deformations by a phase \cite{Huang:2019cja,Moynihan:2020gxj,Emond:2020lwi,Emond:2021lfy}, and the dyon amplitude is therefore generated by the replacement
\[
\cl{M}_3[q,q',p^h] ~~\longrightarrow~~ \cl{M}_3[q,q',p^h]e^{ih\theta}.
\]
Enacting the duality transformation and performing the $q$ and $q'$ integrals, integrating against the wavefunction\footnote{See sections 4.1 and 4.2 in ref. \cite{Kosower:2018adc} for details of how this integration is performed.} as in \cite{Kosower:2018adc}, we find the dyonic shockwave potential
\[
A^\mu_{dyon} 
&= \int \hat{d}^{4} p \frac{1}{p^{2}} \hat{d}^{4}q\hat{d}^{4}q'\hat{\delta}(q^2)\hat{\delta}(q'^2)|\varphi|^2\hat{\delta}^{(4)}(q+q'+p)\\
& \kern+80pt \times \Im\left[\sum_h\varepsilon^{h\mu}(p)\epsilon^{-h\nu}(p)e^{-ih\theta}\cl{M}_{3\nu}[q,q']e^{-ip\cdot x}\right]\\
&= \int \hat{d}^{4} p \frac{1}{p^{2}}\hat{\delta}(2p\cdot q + p^2)\cos(p\cdot x)\left(g_eq^\mu + g_m\frac{\epsilon^\mu(q,p,n)}{p\cdot n}\right)
\]
where we use the notation
\[
	\epsilon^{\mu}(a,b,c) = \epsilon^{\mu\nu\rho\sigma}a_\nu b_\rho c_\sigma,~~~~~\epsilon^{\mu\nu}(b,c) = \epsilon^{\mu\nu\rho\sigma}b_\rho c_\sigma
\]
 and we have defined $g_e = Q\cos\theta, g_m = Q\sin\theta$. In order to express the gauge field in the form above, we have used
\[
\varepsilon^{h\mu}(p)\varepsilon^{-h\nu}(p) = -\eta^{\mu\nu} + \frac{p^\mu n^\nu + p^\nu n^\mu}{p\cdot n}  + ih\frac{\epsilon^{\mu\nu}(p,n)}{p\cdot n},
\]
where $n^\mu$ is a (lightcone) gauge vector with $p\cdot n \neq 0$.
Performing the integrals with the help of Appendix \ref{integralcomp} then gives
\begin{align}
A^\mu_{dyon}(x) &= \frac{g_e}{8\pi}q^\mu\delta(q\cdot x)\log(x^2) \nn\\
&+ \frac{g_m}{2\pi}\delta(q\cdot x)\frac{\epsilon^{\mu\nu\rho\sigma}n_\nu q_{\sigma}x_\rho}{\sqrt{x^2n^2 - (x\cdot n)^2}}\tan^{-1}\left(\frac{x\cdot n}{\sqrt{x^2n^2 - (x\cdot n)^2}}\right).
\end{align}
We can compare this with previous expressions by choosing a null source $q^\mu = (-1,0,0,1)$ and $n^\mu = (0,1,0,0)$, so that the Dirac string is orthogonal to the shockwave, in which case we find
\[
A^\mu_{dyon} = \frac{g_e}{8\pi}q^\mu\delta(t-z)\log|x^2+y^2| - \frac{g_m}{2\pi}q^\mu\delta(t-z)\tan^{-1}\left(\frac{x}{y}\right),
\]
whose magnetic part matches the dual potential given in eq. \eqref{eq:mag_shock}.

%%%%%%%%%%%%%%%%%%%%%%%%%%%%%%%%%%%%%%%%%%%%%%%%%%%%%%%%%%%
\subsection{Dyonic conformal primary wavefunctions}
%%%%%%%%%%%%%%%%%%%%%%%%%%%%%%%%%%%%%%%%%%%%%%%%%%%%%%%%%%%

In general, dyon potentials can be derived by duality rotating purely electric or magnetic potentials, where the rotation mixes the contributions coming from $F_{\mu\nu}$ and its dual. The (anti-)self-dual nature of the conformal primary $A^\Delta_{\mu,a}$ and its shadow are well known \cite{Donnay:2018neh,Pasterski:2020pdk}, and in fact the conformal primary is \emph{locally} an eigenvector of the following self-duality operator                                                                              
\[
D_{a\nu}^\mu = \frac{1}{2}\left(\delta^\mu_{~\nu} - is(a)\frac{\epsilon^\mu_{~\nu}(q,x)}{-q\cdot x}\right),
\]
with eigenvalue $+1$, where $s(z) = -1$ and $s(\bar{z}) = +1$, and where we work without specifying an $i\epsilon$ prescription though this can be straightforwardly included.
This is simply the statement that the conformal primary 
\[
A_{\mu; a}^{\Delta} = \left(\partial_a q_\mu + \frac{\partial_a q \cdot x}{-q \cdot x} q_\mu\right)\frac{1}{(-q \cdot x)^{\Delta}}
\]
can equivalently be expressed in a suitably chosen coordinate patch as
\[
A_{\mu; a}^{\Delta} = is(a) A^{M, \Delta}_{\mu;a}=is(a)\frac{\epsilon_\mu\left(\pd_aq,q,x\right)}{q\cdot x}\frac{1}{(-q \cdot x)^{\Delta}}.
\]
Note that while this is true locally, this is not a global statement, since the vector and its dual cannot be identified under e.g. spatial inversions, due to the pseudo-vector nature of the dual.

Duality rotations are most obviously applied to the self-dual and anti-self-dual projections of the field strengths, i.e.
\[
F_{\mu\nu} \mapsto e^{i\theta}F_{\mu\nu}^+ + e^{-i\theta}F_{\mu\nu}^- = 2\Re(e^{i\theta}F_{\mu\nu}^+).
\]
Thus, for self-dual (anti-self-dual) configurations, $F_{\mu\nu}^- = 0$ ($F_{\mu\nu}^+ = 0$) and the duality transformation is a simple phase deformation leading to a complex coupling.
The spin-1 conformal primary $A_{\mu;a}^{\Delta}$ is either self-dual ($a =\bar{z}$) or anti-self-dual ($a = z)$, and in radial gauge it can be expressed by
\[
A_{\mu;a}^{\Delta} = \frac{1}{\Delta-1}F_{\mu\nu;a}^{\Delta}x^\nu.
\]
 In particular, as the conformal primary is (anti-)self-dual, performing a duality rotation gives the dyon conformal primary
\begin{align}
g {A}_{\mu; a}^{dyon,\Delta} &= ge^{is(a)\theta}A_{\mu; a}^{ \Delta} = (g_e+is(a) g_m) A_{\mu; a}^{\Delta}
\end{align}
 where we have defined $g_e = g\cos\theta$ and $g_m = g\sin\theta$. This can be equivalently expressed as 
 \begin{align}
   g {A}_{\mu; a}^{dyon,\Delta} &= - (g_m-is(a) g_e) A_{\mu; a}^{M\Delta}
 \end{align}
 though it is important to note that this is only true in a specific patch, the relation differing by an overall sign in the other coordinate patch of the celestial sphere. 

%%%%%%%%%%%%%%%%%%%%%%%%%%%%%%%%%%%%%%%%%%%%%%%%%%%%%%%%%%%
\section{Scattering on electric and magnetic backgrounds}
\label{sec:bgscat}
%%%%%%%%%%%%%%%%%%%%%%%%%%%%%%%%%%%%%%%%%%%%%%%%%%%%%%%%%%%

We will consider the scattering of both electrically and magnetically charged scalars in both magnetic and electric backgrounds. To compute the scattering amplitudes we only need classical equations of motion but to be more general we will start with the one-potential Lagrangian formulation of scalar QED \cite{Blagojevic:1983ft}.  The Lagrangian is 
\begin{align}
\mathcal{L}=\mathcal{L}_{\rm Max}+\mathcal{L}_E+\mathcal{L}_M
\end{align}
where $\mathcal{L}_{\rm Max}$ is the usual Maxwell Lagrangian
\begin{align}
\mathcal{L}_{\rm Max}=-\frac{1}{4} F_{\mu\nu}F^{\mu\nu}
\end{align}
but where the field strength is given as 
\begin{align}
\label{eq:1p_FS}
F_{\mu\nu}=\partial_\mu A_\nu-\partial_\nu A_\mu+\epsilon_{\mu \nu \lambda \sigma}G^{\lambda \sigma}~.
\end{align}
The field $G_{\mu\nu}$ is defined in terms of a conserved magnetic current $j^M_\mu$
\begin{align}
G_{\mu \nu}(x)=-\int d^4 y~ h_\mu(x-y) j^M_\nu(y)
\end{align}
where the function $h_\mu(x)$ satisfies
$
\partial^\mu h_\mu(x)=-\delta^{(4)}(x)~
$.
The key aspect of this definition is that as $\partial^\mu(G_{\mu\nu}-G_{\nu\mu})=j^M_\nu$ one has 
\begin{equation}
    \partial_\mu *{F}^{\mu}{}_\nu=-j_\nu^M~
\end{equation}
so that the equations include magnetically charged particles. 
For a theory of magnetically charged complex scalars $\Phi^M$ the magnetic current is
\begin{align}
j_\mu^M=\frac{4\pi}{e^2}\left(-i\gm {\Phi}^M{}^\ast \partial_\mu \Phi^M+i \gm \partial_\mu{\Phi}^M{}^\ast\Phi^M+2\gm^2 \Bm_\mu |\Phi^M|^2\right)
\end{align}
with
\begin{align}
\Bm_\mu(x)=\frac{2\pi }{e^2}\int d^4 y~ \epsilon_{\mu \lambda\alpha\beta}h^\lambda (y-x) F^{\alpha\beta}(y)~.
\end{align}
 It is important to note that in this expression $\Bm_\mu$ is defined implicitly as it appears on both the left and right side through $F^{\alpha\beta}$. This relation can be expanded in powers of the coupling $\gm$ and at leading order we have that 
 \begin{align}
 \bar{\Bm}^\mu(p)=\frac{4\pi}{e^2} \bar{A}^{M\mu}(p)=\frac{4\pi}{e^2}\epsilon^{\mu\lambda\alpha\beta} \frac{n_\lambda p_\alpha \bar{A}_\beta(p)}{n\cdot p}
 \end{align}
 that is, the field $\Bm_\mu$ is just the magnetic dual of $A_\mu$ \eqref{eq:dual_pot}.  Magnetically charged scalar fields are coupled in a non-minimal fashion 
\begin{align}
\mathcal{L}_M=-\partial^\mu {\Phi}^M{}^\ast\partial_\mu \Phi^M+\gm^2 \Bm^2 |\Phi^M|^2~
\end{align}
while electrically charged scalars $\Phi^E$ are coupled in the usual way
\begin{align}
\mathcal{L}_E=-{D}^\ast_\mu {\Phi}^E{}^\ast D^\mu \Phi^E
\end{align}
where the covariant derivative is $D_\mu =\partial_\mu-ig_e A_\mu$.
The resulting equations of motion are 
\begin{align}
\partial^\alpha F_{\alpha \beta}=-j^E_\beta~,
\end{align}
where 
\begin{align}
~~~j^E_\beta=-i g_e {\Phi}^E{}^\ast\partial_\beta \Phi^E+i g_e \partial_\beta  {\Phi}^E{}^\ast \Phi^E-2g_e^2 A_\beta |\Phi^E|^2~.
\end{align}
For the purpose of computing the tree-level scattering, the equation of motion for $\Phi^E$ is the usual 
\begin{align}
\label{eq:el_scalar}
\Box \Phi^E-2ig_e A^\mu \partial_\mu \Phi^E-ig_e(\partial^\mu A_\mu)\Phi^E-g_e^2 A^2 \Phi^E=0
\end{align}
and while the computation of the equation of motion for $\Phi^B$ is slightly more involved, the final result has the same form but with the magnetic coupling $g_e\to -\gm$ and the potential $A_\mu \to \Bm_\mu$
\begin{align}
\label{eq:mag_scalar}
\Box \Phi^M+2i\gm \Bm^{\mu} \partial_\mu \Phi^M+i\gm(\partial^\mu \Bm_\mu)\Phi^M-\gm^2 \Bm^2 \Phi^M=0~.
\end{align}
While the expression for $\Bm_\mu$ is non-linear in the fields, at lowest order in $\gm$ the equation simply describes a scalar moving in the dual potential background. 

\paragraph{Electrically charged scalar}
We start with the equation of motion for the electrically charged scalar field. To compute the scattering we add a source current $J^E$ so that the equation of motion \eqref{eq:el_scalar} in momentum space is to $\mathcal{O}(g_e)$
	\begin{align}
	\label{eq:BB_em}
	-p^2 \bar{\Phi}^E(p)+g_e\int \frac{d^4p'}{(2\pi)^4}  (p+p')^\mu\bar{A}_\mu(p-p') \bar{\Phi}^E(p')
	&=\bar{J}^E(p)~.
	\end{align}
	 At leading order this is simply solved by $\bar \Phi^{E,(0)}(p)=-\frac{\bar J^E(p)}{p^2}$ while at $\mathcal{O}(g_e)$ we have the correction
	\begin{equation}
	\bar \Phi^{E,(1)}(p)=\frac{g_e}{p^2} \int\frac{d^4p'}{(2\pi)^4}\bar   A^\mu(p-p')  (p_\mu+p'_\mu)\bar \Phi^{E,(0)}(p').
	\end{equation}
	The tree-level two-point scattering amplitude is given by the functional derivative of this classical solution, or explicitly, 
	\begin{align}
	\mathcal A^E_2(p_1,p_2)&=-\lim_{p_1^2\to 0}\lim_{p_2^2\to 0} p_1^2p_2^2 \frac{\delta\bar \Phi^E(-p_2)}{\delta \bar J^E(p_1)}
	\end{align}
	and at leading order in $g_e$, we have that the scattering amplitude is 
	\begin{equation}
	\label{eq:l1bbamp}
	\mathcal{A}^{E,(1)}_2(p_1,p_2)=g_e (p_1-p_2)\cdot \bar{A}(-p_1-p_2)~,
	\end{equation}
	where here the massless momenta are on-shell, $p_1^2=p_2^2=0$. In the following we will drop the superscript $(1)$ and the subscript $2$ to avoid notational clutter.

\paragraph{Magnetically charged scalar}
Now we consider a complex, magnetically charged scalar field.  In order to compute the scattering we again add a source term $J^M$ and in momentum space we can write the equation of motion to leading order in $\gm$
\begin{align}
\label{eq:MS_em}
-p^2 \bar{\Phi}^M(p)-\gm\int \frac{d^4p'}{(2\pi)^4}  (p+p')^\mu\bar{\Bm}_\mu(p-p') \bar{\Phi}^M(p')=\bar{J}^M(p)~.
\end{align}
This gives for two-point amplitudes to leading order in $\gm$
\begin{align}
\mathcal{A}^{M,(1)}_2
&=-\gm(p_1-p_2)\cdot \bar{\Bm}(-p_1-p_2)\nn\\
&=-\frac{8\pi}{e^2} \gm \frac{\epsilon_{\mu \nu \rho \sigma} p_2^\mu n^\nu p_1^\rho \bar{A}^\sigma(-p_1-p_2)}{n\cdot (p_1+p_2)}~.
\end{align}
Again we will drop the superscript $(1)$ and the subscript $2$ in the following.

We see that this is the same amplitude as that of an electrically charged particle in an S-dual background $A_\mu \to B_\mu=\frac{4\pi}{e^2} A^M_\mu+\mathcal{O}(g_m)$
\begin{align}
\label{eq:mag_2pt}
\mathcal{A}^{M}(A_\mu)=\mathcal{A}^{E}(\tfrac{4\pi}{e^2} A^M_\mu) 
\end{align} 
but where we also replace the coupling $g_e\to -g_m$. This relation turns out to have an interesting holographic interpretation which is revealed when expressing the scattering amplitudes and backgrounds in a conformal primary basis.

%%%%%%%%%%%%%%%%%%%%%%%%%%%%%%%%%%%%%%%%%%%%%%%%%%%%%%%%%%%
\subsection{Conformal primary backgrounds}
%%%%%%%%%%%%%%%%%%%%%%%%%%%%%%%%%%%%%%%%%%%%%%%%%%%%%%%%%%%

Since we are interested in the boundary manifestation of bulk S-duality our primary focus is on scattering on backgrounds that have a boundary interpretation on the celestial sphere at null infinity.
We first consider scalar scattering on backgrounds corresponding to general conformal primary gauge potentials. We parameterise the external momenta as 
\begin{align}
\label{eq:mom_cel}
p^\mu_i=\eta_i \omega_i (1+z_i \bar{z}_i, z_i+\bar{z}_i, -i (z_i -\bar{z}_i), 1-z_i \bar{z}_i)
\end{align}
with $\eta_1=-\eta_2=1$. To define the celestial amplitudes we will compute the Mellin transform in the parameters $\omega_1$ and $\omega_2$ given by
\begin{equation}
   \tilde{\mathcal A}(z_1,\Delta_1,z_2,\Delta_2)\equiv \int_0^\infty d\omega_1 \omega_1^{\Delta_1-1} d\omega_2 \omega_2^{\Delta_2-1}\mathcal A(p_1,p_2)~.
\end{equation}
It will be convenient to introduce the notation
\begin{align}
2\pi\boldsymbol{\delta}(s)\equiv\int_0^\infty d\omega ~ \omega^{s-1}~.
\end{align}
Where the background depends on a null reference vector $q$, we will now choose a parameterisation\footnote{In contrast to \eqref{eq:qmom1}, we have introduced a subscript on the sphere coordinates for clarity.}
\begin{align}
\label{eq:ref_cel}
q^\mu=(1+z_q \bar z_q, z_q+\bar{z}_q, -i (z_q-\bar{z}_q), 1-z_q \bar z_q)~.
\end{align}

\paragraph{Electric primary background}
The two-point amplitude in the conformal primary background \eqref{eq:cpw}, labelled by the null vector~$q$, conformal dimension $\Delta_q$, and polarisation $a$ (which we take to be either $z_q$ or $\bar z_q$), is 
\begin{align}
\mathcal{A}^{E}_{\Delta_q,a}(p_1,p_2)&=g_e(2\pi)^4 (i)^\Delta \frac{\Delta_q-1}{\Delta_q!}\partial_a q\cdot (p_1-p_2)\nn\\
&\kern+40pt \times  \int_0^\infty d\omega_q~ \omega_q^{\Delta_q-1}\delta^{(4)}(p_1+p_2-\omega_q q)e^{-\epsilon \omega_q}~.
\end{align}
For real momenta, the Dirac-$\delta$ constraint can only be satisfied in soft and collinear regions and so the amplitude receives contributions only from these kinematic regions. However, it can be seen that the contribution from each of these regions is in fact zero and so the amplitude vanishes. The soft regions are where $\omega_q=0$ or one of the external scalar momenta is soft. In the latter case, for example if $\omega_1=0$, the amplitude vanishes as $p_2 \| q$ so that $\partial_a q\cdot p_2=0$. Similarly, the contribution from the collinear region, where $p_1\|p_2\|q$ and $\omega_q=\epsilon_1 \omega_1+\epsilon_2 \omega_2$, vanishes. Finally the region $\omega_q=0$ results in the integral vanishing except for $\Delta_q=1$ in which case the prefactor vanishes and so the amplitude vanishing overall. This is consistent with the very similar analysis of \cite{Chang:2022seh}. 

If we instead treat the momenta as complex, and so the complex coordinates $z_i$ and $\bar{z}_i$ in \eqref{eq:mom_cel} as independent, we can find non-trivial solutions. In particular, if we choose to solve for the holomorphic coordinates we can write the momentum conservation Dirac-$\delta$ as 
\begin{align}
\label{eq:3ptdel_hol}
\delta^{(4)}(p_1+p_2-\omega_q q)=\frac{1}{4 \omega_1 \omega_2 \bar{z}_{12}^2} \delta(\omega_1 -\frac{\bar{z}_{2q}}{\bar{z}_{12}}\omega_q)\delta(\omega_2 -\frac{\bar{z}_{1q}}{\bar{z}_{12}}\omega_q)\delta(z_{1q})\delta(z_{2q}).
\end{align}
We then find for the Mellin transformed two-point function in the positive helicity conformal primary background (i.e. for $a=z_q$)
\begin{align}
\label{eq:CPWgenDel}
\tilde{\mathcal{A}}^{E}_{\Delta_q,z_q}= i^{\Delta_q} (2\pi)^5 g_e  \frac{\Delta_q-1}{\Delta_q!} \boldsymbol{\delta}(\Delta_1+\Delta_2+\Delta_q-3)\frac{ {\delta(z_{1q})\delta(z_{2q})}}{ \bar{z}^{\Delta_1+\Delta_2-1}_{12}\bar{z}_{2q}^{1-\Delta_1} \bar{z}_{1q}^{1-\Delta_2}} ,
\end{align}
while  in the negative helicity background  (i.e. for $a=\bar z_q$) we get $\tilde{\mathcal{A}}^{E}_{\Delta_q,\bar{z}_q}=0$. 
If we had instead solved for the anti-holomorphic coordinates, $\bar{z}_1$ and $\bar{z}_2$, we would find  $\tilde{\mathcal{A}}^{E}_{\Delta_q,{z}_q}=0$ and $\tilde{\mathcal{A}}^{E}_{\Delta_q,\bar{z}_q}$ being given by \eqref{eq:CPWgenDel} with $\bar{z}_{i}\to z_{i}$. The chiral form of the two-point amplitudes correlates with the fact that conformal primary backgrounds have definite duality property: The positive helicity solution $A^\Delta_{\mu;z_q}$ is anti-self dual yielding the anti-holomorphic coordinate dependence in the celestial amplitude $\tilde{\mathcal{A}}^{E}_{\Delta_q,z_q}$, while the holomorphic coordinate dependence in $\tilde{\mathcal{A}}^{E}_{\Delta_q,\bar{z}_q}$ is related to the negative helicity solution $A^{\Delta_q}_{\mu;z_q}$ being self dual. 

\paragraph{Magnetic primary background}
For the dualised background given by the pseudo-conformal primary \eqref{eq:dual_mom_cpw}, the electric amplitude is 
\begin{align}
\mathcal{A}^{E}_{M,\Delta_q,a}(p_1,p_2)&=2g_e(2\pi)^4 (i)^{\Delta_q} \frac{\Delta_q-1}{\Delta_q!}\frac{\epsilon_{\mu\nu\rho\sigma}p^\mu_1 n^\nu p_2^\rho\partial_a q^\sigma}{n\cdot(p_1+p_2)}\nn\\
&\kern+40pt \times  \int_0^\infty d\omega_q~ \omega_q^{\Delta_q-1}\delta^{(4)}(p_1+p_2-\omega_q q)e^{-\epsilon \omega_q}~.
\end{align}
While it is not immediately obvious, the $n$ dependence must drop out due to gauge invariance and this can be explicitly checked. 

In fact, using the parameterisation \eqref{eq:mom_cel} and \eqref{eq:3ptdel_hol}, we find the celestial amplitude in the S-dual background
\begin{align}
\tilde{\mathcal{A}}^{E}_{M,\Delta_q,z_q}= i^{\Delta_q-1} (2\pi)^5 g_e  \frac{\Delta_q-1}{\Delta_q!} \boldsymbol{\delta}(\Delta_1+\Delta_2+\Delta_q-3)\frac{\delta(z_{1q})\delta(z_{2q}) }{ \bar{z}^{\Delta_1+\Delta_2-1}_{12}\bar{z}_{2q}^{1-\Delta_1} \bar{z}_{1q}^{1-\Delta_2}}.
\end{align}
\paragraph{Celestial amplitudes: from S-duality to T-duality}
We thus see that the celestial amplitude in the magnetic positive helicity background is simply related to the amplitude in the original electric positive helicity background by
\begin{equation}
\label{eq:td1}
\tilde{\mathcal{A}}^{E}_{M,\Delta_q,z_q}=-i\tilde{\mathcal{A}}^{E}_{\Delta_q,z_q},
\end{equation}
while the amplitudes in the negative helicity background vanish. Alternatively solving the Dirac-$\delta$ for the anti-holomorphic coordinates, we find a simple relation between the magnetic and electric amplitudes in the negative helicity background
\begin{equation}
\label{eq:td2}
\tilde{\mathcal{A}}^{E}_{M,\Delta_q,\bar{z}_q}=i\tilde{\mathcal{A}}^{E}_{\Delta_q,\bar{z}_q},
\end{equation}
while the amplitudes on the positive helicity background both vanish. 

These relations can be combined and, recalling the definition of the magnetic amplitude \eqref{eq:mag_2pt}, we see that 
\begin{align}
\label{eq:tdual_amp}
\tilde{\mathcal{A}}^{M}_{\Delta_q,a}(z_1,\Delta_1,z_2,\Delta_2)=\epsilon_{ab}\left. \tilde{\mathcal{A}}^{E}_{\Delta_q,b}(z_1,\Delta_1,z_2,\Delta_2)\right|_{ g_e=-{g}_m}~, 
\end{align}
where $\epsilon_{ab}$ is the two-dimensional Levi-Civita tensor in the transverse directions with $\epsilon_{12}=1$. Here $a,b=z_q^1,z_q^2$ are real coordinates, and the factor of $i$ in the transformation rules \eqref{eq:td1} and \eqref{eq:td2}
 follows from the use of complex coordinates. Such a relation is suggestive of a two-dimensional T-duality transformation. This in turn suggests the quite pleasing result that four-dimensional S-duality acting on a bulk theory corresponds to a T-duality type transformation in the boundary celestial conformal field theory. However, it is important to note that the T-duality statement here is purely at the classical level and it remains to be seen what becomes of this boundary interpretation of S-duality in the celestial conformal field theory at the quantum level. We will see that this relation holds in a number of backgrounds.

%%%%%%%%%%%%%%%%%%%%%%%%%%%%%%%%%%%%%%%%%%%%%%%%%%%%%%%%%%%
\subsection{Logarithmic background}
%%%%%%%%%%%%%%%%%%%%%%%%%%%%%%%%%%%%%%%%%%%%%%%%%%%%%%%%%%%

A particularly interesting case is the background given by the logarithmic mode \eqref{eq:log}. For the Feynman regulated expression \eqref{eq:AMlogF} the two-point amplitude evaluates to 
\begin{align}
\mathcal{A}^{E}_{log,a}(p_1,p_2)=2g_e (2\pi)^2\frac{(p_1 \cdot q)( p_2 \cdot \partial_a q)-(p_2\cdot q)( p_1\cdot \partial_a q)}{(p_1\cdot p_2)((p_1+p_2)\cdot q)^2}~.
\end{align}
Performing the Mellin transformation,  using the momentum parameterisation \eqref{eq:mom_cel}, and choosing $a=z_q$, we have for the celestial amplitude 
\begin{align}
\tilde{\mathcal{A}}^{E}_{log,z_q}(z_1,\Delta_1, z_2, \Delta_2)
&=-(2\pi)^2 g_e 
\int  d\omega_1d\omega_2 \omega_1^{\Delta_1-1}\omega_2^{\Delta_2-1}
\frac{\left[\frac{1}{\bar z_{1q}}-\frac{1}{\bar z_{2q}} \right]^{-1}}{(\omega_1|z_{1q}|^2-\omega_2|z_{2q}|^2)^2}
\nn\\
&=\frac{ (2\pi)^3 g_e (-1)^{\Delta_1-1} \Gamma(\Delta_1)\Gamma(\Delta_2)\boldsymbol{\delta}(\Delta_1+\Delta_2-2) }{|z_{12}|^{\Delta_1+\Delta_2}|z_{1q}|^{ \Delta_1-\Delta_2}|z_{2q}|^{\Delta_2-\Delta_1}}\Big[\frac{1}{{z}_{1q}}-\frac{1}{{z}_{2q}}\Big]\nn\\
&= \partial_{z_q}\frac{(2\pi)^3 g_e (-1)^{\Delta_1-1} \Gamma(\Delta_1)\Gamma(\Delta_2-1)\boldsymbol{\delta}(\Delta_1+\Delta_2-2)}{|z_{12}|^{\Delta_1+\Delta_2}|z_{1q}|^{ \Delta_1-\Delta_2}|z_{2q}|^{\Delta_2-\Delta_1}}~.
\end{align}
This amplitude takes the form of a descendant of a scalar CFT three-point correlation function  $\langle \mathcal O_{\Delta_1}(z_1) \mathcal O_{\Delta_2}(z_2)\partial_{z_q}\mathcal O_{\Delta'_q} (z_q)\rangle$ with $\Delta'_q=0$. The logarithmic mode, which is itself a conformal primary with $\Delta_q=1$ and $J_q=1$\cite{Donnay:2018neh}, may indeed be expressed as descending from an operator with $\Delta_q'=0$ and $J_q'=0$ \cite{Pasterski:2021fjn}
\begin{equation}
\bar V^{\log,F}_{\mu;a}(p)=\partial_a\left(8\pi^2\log[-x^2]\frac{q_\mu}{(q\cdot p)p^2}\right),
\end{equation}
which up to total derivatives could have been anticipated from \eqref{Vlog}.
Note that this parent operator takes a very similar form to that of the electric shockwave primary solution \cite{Pasterski:2020pdk}, for which $(q\cdot x)^{-1}\mapsto \delta(q\cdot x)$, which in \cite{Gonzo:2022tjm} was identified as a background that gives rise to standard CFT correlation functions.

The above amplitude is similar to that computed in \cite{Chang:2022seh} though we take both external scalars to be conformal primaries as opposed to one being a shadow conformal primary. 
In two dimensions, the shadow of a scalar conformal primary operator $\mathcal{O}_\Delta(z)$ is a scalar conformal primary of dimension $2-\Delta$ defined as 
\begin{equation}
    \mathbb{S}[\mathcal {O}_{\Delta}](\tilde z)=\frac{\Delta-1}{\pi}\int d^2z \frac{1}{|\widetilde z-z|^{2(2-\Delta)}}\mathcal O_\Delta(z)~.
\end{equation}
We can thus find the celestial amplitude with a scalar operator of dimension $\Delta_1$ and a shadow scalar operator of dimension $\tilde \Delta_2=2-\Delta_2$ in the logarithmic mode background by computing 
\begin{align}
  \mathbb{S}_2[\tilde{\mathcal{A}}^{E}_{ log,z_q}](z_1,\Delta_1, \tilde z_2, \tilde\Delta_2)  &=\frac{1-\tilde \Delta_2}{\pi } \int d^2z_2 \frac{\tilde{\mathcal{A}}^{E}_{log,z_q}(z_1,\Delta_1, z_2, 2-\tilde\Delta_2)}{|\tilde z_2-z_2|^{2\tilde\Delta_2}} ~.
\end{align}
The integral is divergent and needs to be regulated which we do by setting $\Delta_1-\tilde\Delta_2=2\mu$ and then taking the $\mu\to 0$ limit. A similar phenomenon was seen in the computation of \cite{Chang:2022seh} where the massless limit of massive conformal primaries was used to compute the amplitude. The result is (see appendix \ref{app:shadow})
\begin{align}
 \mathbb{S}_2[ \tilde{\mathcal{A}}^{E}_{log,z_q}](z_1,\Delta_1, \tilde z_2, \tilde \Delta_2)  &=
  -(2\pi)^3 g_e (-1)^{\Delta_1}\Gamma(\Delta_1)\Gamma(2-\tilde \Delta_2)\boldsymbol{\delta}(\Delta_1-\tilde \Delta_2)\nn\\
  & \quad \times \partial_{z_q} \left(\lim_{\mu\to 0}
  \frac{\Gamma(-\mu)}{|z_{1\tilde 2}|^{2\tilde \Delta_2}|z_{1q}|^{2\mu}|z_{\tilde2q}|^{-2\mu}}
  \right)\nn\\
  &=\frac{(2\pi)^4 g_e  (-1)^{\Delta_1}(\tilde \Delta_2-1)\boldsymbol{\delta}(\Delta_1-\tilde \Delta_2)}{2\sin (\tilde \Delta_2 \pi)|z_{1\tilde{2}}|^{2\tilde\Delta_2}}\Big[\frac{1}{{z}_{1q}}-\frac{1}{{z}_{\tilde{2}q}}\Big]~,
\end{align}
which has the same coordinate dependence as \cite{Chang:2022seh} but a different normalisation. 
 
For the dualised logarithmic background \eqref{eq:AMlogF} the celestial amplitude does not take a very illuminating form as the dependence on the Dirac string does not obviously drop out at the perturbative amplitude level -- more on this below. However, for the gauge choice $n^\mu=q^\mu$ the two-point function on the logarithmic background \eqref{eq:AMlogFgf} with $a=z_q$ becomes
\begin{equation}
\tilde{\mathcal{A}}^{E}_{M,log,z_q}=-i\tilde{\mathcal{A}}^{E}_{log,z_q},
\end{equation}
and with an extra sign for $a=\bar z_q$. Thus,  we find that scattering on electric and magnetic logarithmic backgrounds are related by a T-duality type relation.

%%%%%%%%%%%%%%%%%%%%%%%%%%%%%%%%%%%%%%%%%%%%%%%%%%%%%%%%%%%
\subsection{Conformally soft background}
\label{sec:CSscattering}
%%%%%%%%%%%%%%%%%%%%%%%%%%%%%%%%%%%%%%%%%%%%%%%%%%%%%%%%%%%

The amplitude for the conformally soft mode \eqref{eq:ACS} can be split into two parts. The first is
\begin{align}
\mathcal{A}^{E}_{CS',a}(p_1,p_2)=-8\pi^2 g_e \delta(p_1\cdot p_2) \frac{(p_1 \cdot q)( p_2 \cdot \partial_a q)-(p_2\cdot q)( p_1\cdot \partial_a q)}{((p_1+p_2)\cdot q)^2}~,
\end{align}
and corresponds to scattering on the shift descendant background \eqref{eq:ACS'}.
Again using the momentum parameterisation \eqref{eq:mom_cel}, and choosing $a=z_q$, after the Mellin transform we have for the celestial amplitude
\begin{align}
\label{CelestialCS'}
\tilde{\mathcal{A}}^{E}_{CS', z_q}(\Delta_1, \Delta_2)
&=\frac{(2\pi)^3  g_e  (-1)^{\Delta_1} \Gamma(\Delta_1)\Gamma(\Delta_2) \boldsymbol{\delta}(\Delta_1+\Delta_2-2) {\delta}(|z_{12}|^2)}{|z_{1q}|^{\Delta_1-\Delta_2}|z_{2q}|^{\Delta_2-\Delta_1}}\Big[\frac{1}{z_{1q}}-\frac{1}{z_{2q}}\Big]\nn\\
&=\partial_{z_q}\left[\frac{(2\pi)^3  g_e   \frac{(-1)^{\Delta_1}\Gamma(\Delta_1)\Gamma(\Delta_2) }{\Delta_1-\Delta_2}\boldsymbol{\delta}(\Delta_1+\Delta_2-2) {\delta}(|z_{12}|^2)}{|z_{1q}|^{\Delta_1-\Delta_2}|z_{2q}|^{\Delta_2-\Delta_1}}\right].
\end{align}
The second contribution, corresponding to scattering on the electric shockwave descendant background  \eqref{eq:ACS''}, is given by
\begin{align}
\mathcal{A}^{E}_{CS'',a}(p_1,p_2)=8\pi^2 g_e \delta'((p_1+ p_2)\cdot q) \frac{(p_1 \cdot q)( p_2 \cdot \partial_a q)-(p_2\cdot q)( p_1\cdot \partial_a q)}{p_1\cdot p_2}~,
\end{align}
which, for the choice $a=z_q$, gives the celestial amplitude
\begin{align}
\label{CelestialCS''}
\tilde{\mathcal{A}}^{E}_{CS'', z_q}(\Delta_1, \Delta_2)
&=\frac{ 4\pi^3  g_e (\Delta_2-\Delta_1) \boldsymbol{\delta}(\Delta_1+\Delta_2-2)}{|z_{12}|^{\Delta_1+\Delta_2}|z_{1q}|^{\Delta_1-\Delta_2}|z_{2q}|^{\Delta_2-\Delta_1}}\Big[\frac{1}{z_{1q}}-\frac{1}{z_{2q}}\Big]\nn\\
&=\partial_{z_q} \Big[\frac{(2\pi)^3  g_e  \boldsymbol{\delta}(\Delta_1+\Delta_2-2)}{|z_{12}|^{\Delta_1+\Delta_2}|z_{1q}|^{\Delta_1-\Delta_2}|z_{2q}|^{\Delta_2-\Delta_1}}\Big]~.
\end{align}

One can clearly see that the contributions from the two parts of the conformally soft mode are closely related with both terms giving a singular contribution in the collinear limit. Moreover, the $CS''$ amplitude is very similar to that computed directly from the logarithmic mode which is to be expected as that amplitude had been shown to have the correct form for a current-scalar-scalar three-point function \cite{Chang:2022seh}. As discussed in section \ref{CSmodes}, the conformally soft backgrounds can be viewed as descending from operators with $(\Delta_q',J_q')=(0,0)$. The $CS''$ background descends from the electric shockwave solution \eqref{shiftshockFourier}, repeated here for convenience,
\begin{equation}
    \bar A^{CS''}_{\mu;a}
    =\partial_a \bar A^{E-shock}_{\mu}\quad\text{with}\quad 
    \bar A^{E-shock}_{\mu}=8\pi^2q_\mu\frac{\delta(q\cdot p)}{p^2}.
\end{equation}
Scattering two scalars on the latter gives rise to a standard CFT three-point function \cite{Gonzo:2022tjm}, 
\begin{equation}
\langle \mathcal O_{\Delta_1}(z_1)\mathcal O_{\Delta_2}(z_2)\mathcal O^{E-shock}_{\Delta_q'}(z_q)\rangle = \frac{(2\pi)^3 g_e \boldsymbol{\delta}(\Delta_1+\Delta_2+\Delta_{q}'-2) }{|z_{12}|^{\Delta_1+\Delta_2-\Delta_q'} |z_{2q}|^{\Delta_2+\Delta_q'-\Delta_1} |z_{q1}|^{\Delta_q'+\Delta_1-\Delta_2}},
\end{equation}
and in \eqref{CelestialCS''} we find its descendant. 
The $CS'$ background can be expressed as descending from the electric shift background \eqref{shiftshockFourier}, repeated here for convenience,
\begin{equation}
 \bar A^{CS'}_{\mu;a}
    =\partial_a \bar A^{E-shift}_{\mu} \quad \text{with} \quad
\bar A^{E-shift}_\mu=-8\pi^2 q_\mu \frac{\delta(p^2)}{q\cdot p}.
\end{equation}
The amplitude for two scalars scattering on the latter takes the form of a non-standard CFT three-point function\footnote{Strictly speaking the shift background correlator contains an ambiguity which is, however, in the kernel of $\partial_a$ for an appropriate choice of regularisation of the Mellin integral and so we drop it here.}
\begin{align}
\langle \mathcal O_{\Delta_1}(z_1)\mathcal O_{\Delta_2}(z_2)\mathcal O^{E-shift}_{\Delta_q'}(z_q)\rangle &= \nn  \\
&\kern-40pt\frac{(2\pi)^3  g_e   \frac{(-1)^{\Delta_1}\Gamma(\Delta_1)\Gamma(\Delta_2) }{\Delta_1-\Delta_2}\boldsymbol{\delta}(\Delta_1+\Delta_2+\Delta_q'-2) {\delta}(|z_{12}|^{\Delta_1+\Delta_2-\Delta_q'})}{|z_{1q}|^{\Delta_1+\Delta_q'-\Delta_2}|z_{2q}|^{\Delta_2+\Delta_q'-\Delta_1}}
\end{align}
whose descendant amplitude is \eqref{CelestialCS'}.

We can similarly compute the amplitudes on the dualised conformally soft backgrounds \eqref{eq:mag_csp} and \eqref{eq:mag_cspp}. For the shift descendant contribution \eqref{eq:ACS'}, the result is again related by two-dimensional T-duality
\begin{align}
\tilde{\mathcal{A}}^{E}_{M,CS',a}=-\epsilon_{ab} \tilde{\mathcal{A}}^{E}_{CS',b}~.
\end{align}
We can also express this as a relation between the electric and magnetic amplitudes,
\begin{align}
\tilde{\mathcal{A}}^{M}_{CS',a}=-\frac{4\pi}{e^2}\epsilon_{ab}\left. \tilde{\mathcal{A}}^{E}_{CS',b}\right|_{g_e\to -g_m}~.
\end{align}
Compared to the case for general primary backgrounds, the result here is somewhat cleaner as neither amplitude is generically vanishing. 
For the shockwave descendant contribution \eqref{eq:ACS'}, there is a dependence on the reference vector $n$, due to the the nature of the Fourier transform, such that the result does not follow this T-dual form. However, if we make the choice $n_\mu=q_\mu$,  such that the dualisation acts by the substitution \eqref{eq:sub} in the dual potential, then we do find the same relation between electric and magnetic amplitudes.

%\paragraph{Conformally soft factorisation}
As in the logarithmic mode background we can consider the celestial amplitudes with a scalar of dimension $\Delta_1$ and a shadow scalar of dimension $\tilde \Delta_2=2-\Delta_2$. 
In the $CS''$ background we find
\begin{align}
\label{S2CS''}
     \mathbb{S}_2[ \tilde{\mathcal{A}}^{E}_{CS'',z_q}](z_1,\Delta_1, \tilde z_2, \tilde \Delta_2)
     &=\frac{1-\tilde{\Delta}_2}{\pi}\partial_{z_q} \int d^2z_2 \frac{(2\pi)^3 g_e \boldsymbol{\delta}(\Delta_1-\tilde \Delta_2)}{|z_{12}|^{\Delta_1-\tilde\Delta_2+2}|z_{1q}|^{\Delta_1+\tilde\Delta_2-2}|z_{2q}|^{2-\Delta_1-\tilde\Delta_2}{|z_{\tilde2 2}|^{2\tilde \Delta_2}}}\nn\\
     &=\frac{\Gamma(2-\tilde \Delta_2)\Gamma\left(\frac{\Delta_1+\tilde \Delta_2}{2}\right)\Gamma\left(\frac{\tilde \Delta_2-\Delta_1}{2}\right)}{\Gamma(\tilde \Delta_2)\Gamma\left(1-\frac{\Delta_1+\tilde \Delta_2}{2}\right)\Gamma\left(1+\frac{\Delta_1-\tilde \Delta_2}{2}\right)}\times\nn\\
     &\quad\partial_{z_q} \frac{(2\pi)^3  g_e\boldsymbol{\delta}(\Delta_1-\tilde\Delta_2)}{|z_{1\tilde2}|^{\Delta_1+\tilde\Delta_2}|z_{1q}|^{\Delta_1-\tilde \Delta_2}|z_{\tilde 2q}|^{\tilde \Delta_2- \Delta_1}}.
\end{align}
Again, we recognise this as the descendant of a standard CFT three-point function
\begin{equation}
\langle \mathcal O_{\Delta_1}(z_1)\mathbb{S}[\mathcal O]_{\tilde \Delta_2}(\tilde z_2)\mathcal O^{E-shock}_{\Delta_q'}(z_q)\rangle = \mathcal{N} \times \frac{(2\pi)^3 g_e  \boldsymbol{\delta}(\Delta_1-\tilde \Delta_2+\Delta_{q}') }{|z_{1\tilde 2}|^{\Delta_1+\tilde \Delta_2-\Delta_q'} |z_{\tilde 2q}|^{\tilde \Delta_2+\Delta_q'-\Delta_1} |z_{q1}|^{\Delta_q'+\Delta_1-\tilde \Delta_2}}.
\end{equation}
If we enforce $\Delta_1=\tilde \Delta_2$ (since $\Delta_q'=0$) then this degenerates to a two-point function $1/|z_{1\tilde 2}|^{\Delta_1+\tilde \Delta_2}$ but with diverging normalisation $\mathcal{N}$. Thus, the descendant correlator \eqref{S2CS''} appears ill-defined.
Meanwhile, computing the $\partial_{z_q}$ derivative before enforcing $\Delta_1=\tilde \Delta_2$ gives an additional factor $(\Delta_1-\tilde \Delta_2)$ which cancels the divergent normalisation $\mathcal{N}$ and instead yields 
\begin{equation}
    \mathbb{S}_2[ \tilde{\mathcal{A}}^{E}_{CS'',z_q}](z_1,\Delta_1, \tilde z_2, \tilde \Delta_2) 
    =-\frac{(2\pi)^3 g_e \boldsymbol{\delta}(\Delta_1-\tilde\Delta_2)(1-\tilde{\Delta}_2)}{|z_{1\tilde2}|^{\Delta_1+\tilde\Delta_2}}\Big[\frac{1}{{z}_{1q}}-\frac{1}{{z}_{\tilde2q}}\Big].
\end{equation}
This takes the form of a (conformally) soft factorisation of a three-point function
\begin{equation}
    \langle J^{CS''}(z_q)\mathcal{O}_{\Delta_1}(z_1) \mathbb{S}[\mathcal O]_{\tilde \Delta_2}(\tilde z_2)\rangle= S^{(0)}(z_1,\tilde z_2,z_q)\langle \mathcal O_{\Delta_1}(z_1)\mathbb{S} [\mathcal O]_{\tilde\Delta_2}(\tilde z_2)\rangle
\end{equation}
into a standard CFT two-point function \footnote{Up to a normalisation this corresponds to the shadow inner product advocated for in \cite{Crawley:2021ivb}.}
\begin{equation}
    \langle \mathcal O_{\Delta_1}(z_1)\mathbb{S} [\mathcal O]_{\tilde\Delta_2}(\tilde z_2)\rangle=\frac{\tilde{\mathcal{N}}\boldsymbol{\delta}(\Delta_1-\tilde\Delta_2)}{|z_{12}|^{\Delta_1+\tilde \Delta_2}}
\end{equation}
albeit with a funny normalisation $\tilde{\mathcal{N}}=\frac{1}{2}(2-\Delta_1-\tilde \Delta_2)$,
multiplied by a (conformally) soft factor (recall that $\eta_{1}=-\eta_2=1$)
\begin{equation}
   S^{(0)}(z_1,\tilde z_2,z_q)= (2\pi)^3 g_e\left(\frac{1}{{z}_{1q}}-\frac{1}{{z}_{\tilde2 q}}\right)=
   (2\pi)^3g_e\frac{z_{1\tilde2}}{z_{\tilde2 q}z_{q1}}.
\end{equation}
We interpret this as the Ward identity of a conformally soft current $J^{CS''}$ with $\Delta_q=1$  that is conserved as $\partial_{\bar z} J^{CS''}(z_q)=0$ up to contact terms. This factorisation of a three-point correlator into a soft prefactor and a two-point function is naturally reminiscent of soft factorisation in flat space amplitudes with the notable distinction that for momentum space amplitudes there is no non-trivial two-point function and so such $3\to2$ factorisation does not occur. 

In the $CS'$ background we have
\begin{align}
\mathbb{S}_2[\tilde{\mathcal{A}}^{E}_{CS',z_q}](z_1,\Delta_1, \tilde z_2, \tilde \Delta_2) &= 
     4\pi^2 g_e   (-1)^{\Delta_1+1} \Gamma(\Delta_1)\Gamma(2-\tilde \Delta_2)\boldsymbol{\delta}(\Delta_1-\tilde \Delta_2)\nn\\
     &\quad \times \partial_{z_q} \int d^2z_2 \frac{\delta(|z_{12}|^2)}{|z_{1q}|^{\Delta_1+\tilde \Delta_2-2}|z_{2q}|^{2-\Delta_1-\tilde\Delta_2}|z_{\tilde 2 2}|^{2\tilde \Delta_2}},
\end{align}
where, as above, we are actually computing the shadow transform of the amplitude in the electric shift background and afterwards descending the latter to the $CS'$ background.
For a particular regularisation of the Dirac-$\delta$ in the shadow integral we find
\begin{equation}
    \int d^2z_2 \frac{\delta(|z_{12}|^2)}{|z_{2q}|^{2-\Delta_1-\tilde \Delta_2}|z_{\tilde2 2}|^{2\tilde \Delta_2}}=\frac{\pi}{|z_{1q}|^{2-\Delta_1-\tilde \Delta_2}|z_{1\tilde 2}|^{2\tilde \Delta_2}}.
\end{equation}
Using this result the shadow transformed amplitude on the shift background becomes
\begin{equation}
\langle \mathcal O_{\Delta_1}(z_1)\mathbb{S}[\mathcal O]_{\tilde \Delta_2}(\tilde z_2)\mathcal O^{E-shift}_{\Delta_q'}(z_q)\rangle =  
4\pi^3 g_e (-1)^{\Delta_1+1} \Gamma(\Delta_1)\Gamma(2-\tilde \Delta_2)
   \frac{\boldsymbol{\delta}(\Delta_1-\tilde \Delta_2)} {|z_{1\tilde 2}|^{2\tilde \Delta_2}}~,
\end{equation}
which up to an awkward normalisation corresponds to a CFT two-point function.
Because of this, the $\partial_{z_q}$ descendant amplitude now vanishes
\begin{equation}
\mathbb{S}_2[\tilde{\mathcal{A}}^{E}_{CS',z_q}](z_1,\Delta_1, \tilde z_2, \tilde \Delta_2)
=0.
\end{equation}

%%%%%%%%%%%%%%%%%%%%%%%%%%%%%%%%%%%%%%%%%%%%%%%%%%%%%%%%%%%
\subsection{Magnetic shockwave}
\label{subsec:BM}
%%%%%%%%%%%%%%%%%%%%%%%%%%%%%%%%%%%%%%%%%%%%%%%%%%%%%%%%%%%

As a further example let us consider an electrically charged scalar in the boosted Dirac monopole background, i.e. the magnetic shockwave solution \eqref{eq:mag_shock}. Using the dualised potential \eqref{eq:dual_emshock}, we find the two-point function
\begin{align}\label{A2Bshock}
\mathcal{A}_{M-shock}^{E}(p_1,p_2)
&=(4\pi)^2  g_e \delta((p_1+p_2)\cdot q )\frac{\epsilon^{\lambda \kappa \mu \nu} n_\kappa p_{1\mu} q_\nu p_{2\lambda} }{(p_1\cdot p_2) (n\cdot (p_1+p_2))}~.
\end{align}
If we had instead used the Fourier transform of the single copy of the NUT-wave we would have essentially the same result. In either case the amplitude depends on the position of the Dirac string. As we will discuss below, this is likely a feature of our perturbative approach and would not appear in a non-perturbative, re-summed amplitude. 

Keeping this in mind we proceed to evaluate the corresponding celestial amplitude 
on the magnetic shockwave background by carrying out a Mellin transform over the external scalar particle energies $\omega_1$ and $\omega_2$. We use the parameterisation \eqref{eq:mom_cel} for the scattered momenta $p_1$ and $p_2$, and \eqref{eq:ref_cel} for the reference vector $q$. A natural and simple choice for the location of the Dirac string is to consider it along a null direction 
\begin{align}\label{nnull}
n^\mu=(1+|z_n|^2, z_n+\bar{z}_n, -i (z_n-\bar{z}_n), 1-|z_n|^2)~.
\end{align}
The result for the celestial amplitude on the magnetic shockwave backround is  
\begin{align}
\tilde{\mathcal{A}}^{E}_{M-shock}(\Delta_1, \Delta_2)
&=
\frac{ (2\pi)^3 g_e  \boldsymbol{\delta}(\Delta_1+\Delta_2-2)\, \Xi(z_1, z_2, z_q, z_n)}{|z_{12}|^{2}|z_{1q}|^{2\Delta_1-2}|z_{2q}|^{2-2\Delta_1}}.
\end{align}

Noting that the magnetic shockwave background represents the dualisation of a conformal primary with $\Delta=0$, we can rewrite this in a suggestive form 
\begin{align}
\tilde{\mathcal{A}}^{E}_{M-shock}(\Delta_1, \Delta_2, \Delta=0)=
\frac{(2\pi)^3 g_e   \boldsymbol{\delta}(\Delta_1+\Delta_2-2) \,\Xi(z_1, z_2, z_q, z_n)}{|z_{12}|^{\Delta_1+\Delta_2-\Delta}|z_{1q}|^{\Delta_1+\Delta_q-\Delta_2}|z_{2c}|^{\Delta_2+\Delta-\Delta_1}}~.
\end{align}
Here we see the structure of a three-point function previously observed for the electromagnetic shockwave in \cite{Gonzo:2022tjm} but, due to the dualisation, modified by the $\Xi$ function which takes the form
\begin{align}
\Xi(z_1, z_2, z_q, z_n)=i\frac{r-\bar{r}}{1-|r|^2}
\end{align}
with the conformal cross ratio
$r=\frac{z_{1n}z_{2q}}{z_{2n}z_{1q}}$. Note that while \eqref{nnull} defines a fourth point on the celestial sphere, the celestial amplitude still takes the form of a three-point function, albeit multiplied by the $\Xi$ function which explicitly depends on $n$.

The fact that the scattering amplitude depends on the position of the Dirac string might seem surprising as it is supposed to be a gauge artefact. However, the explicit dependence, and the corresponding breaking of Lorenz and gauge invariance, in perturbative amplitudes has been previously noted e.g. \cite{Rabl:1969gx, Terning:2018udc}. The reason is that the Dirac string is unobservable when the charge quantisation condition $ g_e {g}_m =2\pi N$ with $N\in \mathbbm{Z}$ is satisfied \cite{Dirac:1931kp}, see \cite{Shnir:2005vvi} for a review of all things monopole. However, perturbative approaches require that  $g_e {g}_m\ll 1$ and so it is not possible to satisfy the quantisation condition and the dependence on the direction $n$ of the Dirac string should only vanish after an all-order re-summation.

%%%%%%%%%%%%%%%%%%%%%%%%%%%%%%%%%%%%%%%%%%%%%%%%%%%%%%%%%%%
\section{Eikonal re-summation}
\label{sec:eik}
%%%%%%%%%%%%%%%%%%%%%%%%%%%%%%%%%%%%%%%%%%%%%%%%%%%%%%%%%%%

In the previous section
we found that for the scattering of an electrically charged scalar in a boosted monopole background there was explicit dependence on the position of the Dirac string. This dependence is likely a feature of the perturbative approach and should vanish when the all-order perturbative series is re-summed. Such a re-summation can be carried out for the IR divergent part of the amplitude \cite{Terning:2018udc} or, equivalently, in the eikonal approximation.  

%%%%%%%%%%%%%%%%%%%%%%%%%%%%%%%%%%%%%%%%%%%%%%%%%%%%%%%%%%%
\subsection{Electrically charged scalar: Wilson loop} 
%%%%%%%%%%%%%%%%%%%%%%%%%%%%%%%%%%%%%%%%%%%%%%%%%%%%%%%%%%%

In the eikonal limit we can re-sum the contributions of the background, so that the connected part of the two-point amplitude in a background potential $A_\mu$ is given by 
\begin{align}
\mathcal{A}^{E, IR}(p_1,p_2)=\text{exp}\Big[g_e \int \frac{d^4 k}{(2\pi)^2} \frac{\bar{A}(-k)\cdot p_2}{k\cdot p_2}\Big]\mathcal{A}^{E}(p_1,p_2)~.
\end{align}
This expression captures an infinite number of interactions of the probe particles with the background but assumes that the momentum transferred to the probe is small. 
This expression can be written as 
$
\mathcal{A}^{E, IR}(p_1,p_2)=\mathcal{W}^E_{{p_2}}\mathcal{A}^{E}(p_1,p_2)
$
with the Wilson line being given by 
\begin{align}
\mathcal{W}^E_{{p_2}}=\text{exp}\Big[i g_e \int ds~ p_2 \cdot A(s p_2)]~.
\end{align}
If we compactify space-time so that a massless particle trajectory can be identified as a closed loop $\mathcal{C}$, we have by Stokes' theorem
\begin{align}
\mathcal{W}^E_{\mathcal{C}} &= \text{exp}\Big[i g_e\int_\mathcal{C} A  \Big]
=\text{exp}\Big[i g_e\int_S ~ dA  \Big]
\end{align}
where $S$ is any surface bounded by $\mathcal{C}$. 

We can consider this expression for the case of a magnetic dual background, where the dual potential explicitly depends on the reference vector $n$. As we have seen, this is closely related to the scattering of electric and magnetic particles and so the Wilson loop can be shown to be independent of the vector $n$ by essentially repeating the argument used for soft-resummation factors in \cite{Terning:2018udc}. Using the definition of the  potential, \eqref{eq:1p_FS}, we can rewrite the field strength as
\begin{align}
\partial_\mu A_\nu(x)-\partial_\nu A_\mu(x)= {F}_{\mu\nu}(x) +\epsilon_{\mu \nu \kappa \lambda}\int d^4y~ h^\kappa (x-y)   j^{M\lambda}(x)~.
\end{align}
Thus we can see that the Wilson loop can be expressed as the  product of two terms.   
The first of which, $\text{exp}\int {F}$, is explicitly independent of the vector $n$. The second, can be rewritten using the identity,
\begin{align}
\int d^4y~ h^\kappa (x-y)   j^{M\lambda}(x) = n^\kappa \int d\tau~  j^{M\lambda}(x-n\tau)~
\end{align}
which follows from the particular choice of $h^\kappa(x-y)$ in section \ref{sec:bgscat} and appropriate boundary conditions.
Hence the Wilson loop can be rewritten as 
\begin{align}
\mathcal{W}_\mathcal{C}^E=\text{exp}\Big[i g_e\int F+  i  \frac{4\pi g_e g_m }{e^2} L \Big]
\end{align}
with 
\begin{align}
 L = \frac{e^2}{ 4\pi {g}_m } \int_S dx^\mu \wedge dx^\nu \int d\tau  \epsilon_{\mu \nu \kappa \lambda} n^\kappa j^{M\lambda}(x-n\tau)~.
\end{align}
Obviously for cases where $j^\lambda=0$, such as for the conformal partial waves, the dependence on $n$ explicitly drops out. More generally, for example in the presence of magnetically charged scalars, the current is non-vanishing and scales as ${g}_m$. 

For potentials sourced by point particles, $L$ is the topological invariant of \cite{Terning:2018udc}. In this case we have that the current is written as an integral over the world line, $y^\lambda(s)$, of the charged particle
\begin{align}
j^\lambda(x)=\frac{4\pi {g}_m}{e^2} \int ds ~u^\lambda \delta^{(4)}(x-y(s))
\end{align}
with $u^\lambda=\tfrac{dy^\lambda}{ds}$ and where we have taken the charge to be $ {g}_m$ in units of $e^2$ so that
\begin{align}
L= \int_S dx^\mu \wedge dx^\nu  \epsilon_{\mu \nu \kappa \lambda} \int d\tau n^\kappa \int ds~ u^\lambda \delta^{(4)}(x-n\tau - y(s))~.
\end{align}
This integral quantity counts the intersection between the surface $S$ bounded by the particle trajectory, and the world-sheet of the string spanned by $y(s)+n\tau$. If we assume that this quantity is indeed integer valued, then when the charge quantisation condition $4\pi g_e {g}_m/e^2=2\pi N$, $N\in \mathbbm{Z}$, is satisfied the contribution to the Wilson loop is a multiple of $2\pi$ and the dependence on the position of the Dirac string vanishes from the amplitude. 

%%%%%%%%%%%%%%%%%%%%%%%%%%%%%%%%%%%%%%%%%%%%%%%%%%%%%%%%%%%
\subsection{Magnetically charged scalar: 't Hooft loop}
%%%%%%%%%%%%%%%%%%%%%%%%%%%%%%%%%%%%%%%%%%%%%%%%%%%%%%%%%%%

For magnetically charged scalars, we can see from the equation of motion, \eqref{eq:mag_scalar}, that 
considering higher powers of ${g}_m$ in the definition of ${B}_\mu$ does not give corrections to the two-point amplitude as each higher-power of ${g}_m$ multiplies an additional field $\Phi^B$ and so only contributes to higher-point amplitudes. 
There are of course higher-${g}_m$ corrections from the quadratic-in-${B}$ term in the equation of motion and from higher terms in the perturbative expansion of \eqref{eq:MS_em}. Just as for the electric case, these higher-order corrections will be IR-divergent and these divergent contributions can be re-summed in the eikonal approximation. 
The resulting two-particle amplitude has the form
\begin{align}
\mathcal{A}^{M, con, IR}_2(p_1,p_2)=\mathcal{W}_p^M \mathcal{A}_2^{M,(1)}(p_1,p_2)~.
\end{align}
where the prefactor is a 't Hooft line
\begin{align}
\mathcal{W}_p^M &= \text{exp}\Big[-{g}_m\int \frac{d^4 k}{(2\pi)^4} \frac{\bar{B}(-k)\cdot p_2}{k\cdot p_2} \Big]\nn\\
&=\text{exp}\Big[-i {g}_m\int ds ~ p_2\cdot  {B}(s p_2)  \Big]~.
\end{align}
The appearance of the 't Hooft line in the IR divergent part of scattering has been previously noted in \cite{Choi:2019sjs} where it was shown the the magnetic analogue of the Faddeev-Kulish dressing had the form of a 't Hooft line, see also \cite{Blagojevic:1981he, Blagojevic:1981sz, Csaki:2022tvb}. If we consider compactified space-time so that massless particle trajectory can be identified as a closed loop $\mathcal{C}$ we have by Stokes theorem
\begin{align}
\mathcal{W}_p^M &= \text{exp}\Big[-i g_m\int_\mathcal{C} B  \Big]
=\text{exp}\Big[-i g_m\int_S ~ dB  \Big]
\end{align}
where $S$ is any surface bounded by $\mathcal{C}$. 

%%%%%%%%%%%%%%%%%%%%%%%%%%%%%%%%%%%%%%%%%%%%%%%%%%%%%%%%%%%
\subsection{Celestial eikonal limit}
%%%%%%%%%%%%%%%%%%%%%%%%%%%%%%%%%%%%%%%%%%%%%%%%%%%%%%%%%%%

%%%%%%%%%%%%%%%%%%%%%%%%%%%%%%%%%%%%%%%%%%%%%%%

\begin{figure}
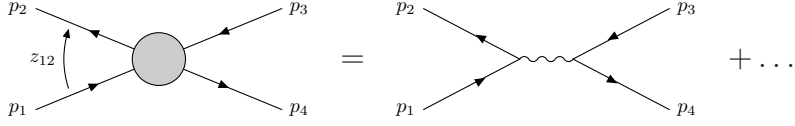

	\centering
$	\vcenter{\hbox{\includegraphics[scale=1]{./Figures/eik_kin.mps}}}$~~~=~~~$\vcenter{\hbox{\includegraphics[scale=1]{./Figures/eik_kin_pert1.mps}}}$~~~$+\dots$
	\caption{Eikonal kinematics corresponds to $|z_{12}| \to 0$. We take all momementa as incoming and the outgoing arrows denote negative energies.}
	\label{fig:eik_kin}
\end{figure}
A version of the  eikonal limit can also be used to make a concrete connection between Mellin transformed four-point amplitudes and the transformed scattering in a background \cite{deGioia:2022fcn}. 
We consider the Mellin transform of the leading-order eikonal amplitude for scalar particles interacting via exchange of a spin-$1$ particle. The four-particle, tree-level amplitude with external electrically charged scalars is
\begin{align}
\mathcal{A}^{EE}_4= i(2\pi)^4g_e^2 \frac{2s+t}{t}\delta^{(4)}(\sum p_i)
\end{align}
where we define $s=-(p_1+p_3)^2$, $t=-(p_1+p_2)^2$. 
We choose the parameterisations for the massless external momenta
\begin{align}
p^\mu_i&=\eta_i \omega_i(1+|z_i|^2, \hat{p}^\perp_i, \iota_i (1-|z_i|^2))
\end{align}
with $\eta_1=\eta_3=-\eta_2=-\eta_4=1$, $\iota_{1}=\iota_2=-\iota_3=-\iota_4=-1$
so that $p_1$, $p_3$ are incoming, $p_2$, $p_4$ are ``outgoing" i.e. incoming with negative energy, and the transverse coordinates are $\hat{p}^\perp_i= (z_i+\bar{z}_i,-i(z_i-\bar{z}_i))$ for $i=1,2,3,4$. In the eikonal limit we take $s\gg -t$, that is 
\begin{equation}
|1+z_1 \bar{z}_3|^2  \omega_3 \gg |z_{12}|^2 \omega_2~.
\end{equation}
This is essentially the same kinematics as in \cite{deGioia:2022fcn} but with $p_2 \leftrightarrow p_3$.
As we wish to consider the Mellin transform in $\omega$-variables, we consider $|z_{12}|\ll 1$ while keeping $\omega_i$ fixed.
In this limit,
the celestial amplitude is given by \cite{deGioia:2022fcn}
\begin{align}
\tilde{\mathcal{A}}^{EE}_{4,eik}= \prod_{i=1}^4 \int \frac{d\omega_i}{\omega_i} \omega_i^{\Delta_i} \mathcal{A}^{EE}_{4,eik}
&=-4(2\pi)^2 i g_e^2
\int {d^2 x_\perp} \int {d^2 \bar{x}_\perp}G_\perp(\vec{x}_\perp, \vec{\bar{x}}_\perp)\nn\\
&\times 
\frac{\Gamma(\Delta_1+\Delta_2)}{(-i \hat{p}^\perp_{12}\cdot \vec{x}_\perp)^{\Delta_1+\Delta_2} }
\frac{\Gamma(\Delta_3+\Delta_4)}{(-i \hat{p}^\perp_{34}\cdot \vec{\bar{x}}_\perp)^{\Delta_3+\Delta_4} }
\end{align}
where 
\begin{equation}
G_\perp(\vec{x}_\perp, \vec{\bar{x}}_\perp)=\int \frac{d^2k_\perp}{(2\pi)^2} \frac{e^{i \vec{k}_\perp\cdot (\vec{x}_\perp-\vec{\bar{x}}_\perp)}}{k_\perp^2}~
\end{equation}
is the transverse propagator and we have used the notation $\hat{p}^\perp_{ij}=\hat{p}^\perp_i-\hat{p}^\perp_j$. We have focussed on the case of single photon exchange, but of course in the eikonal limit it is possible to re-sum an infinite number of contributions. Effectively, this is done by replacing the transverse propagator: $ig_e^2 G_\perp\to \text{exp}(ig_e^2 G_\perp)-1$.

We now consider the general amplitude for an electrically charged scalar scattering in a potential of the form 
\begin{equation}
A^E_\mu(x)=q_{s,\mu} \delta(x^-) a^E(\vec{x}_\perp)
\end{equation} 
where we choose coordinates $x^\mu=(x^+,x^-,\vec{x}_\perp)$ and take the shockwave momentum to be $q_{s,\mu}=(-1,0,0,1)$.
Working to leading order in $g_e$, we have that the two-particle scattering amplitude is 
\begin{equation}
\mathcal{A}_{2;E}^{E}=-4g_e \pi p^-_1 \delta(p_1^-+p_2^-)\bar{a}^E(-\vec{p}^{\perp}_1-\vec{p}^{\perp}_{2})~,
\end{equation}
where we have defined $p^-=-2p\cdot q_s$. 
Now we use the same parameterisation as above such that, $p_1^-=\omega_1$,  $p_2^-=-\omega_2$ and $\vec{p}^\perp_i=\omega_i \eta_i\hat{p}^\perp_i$ for $i=1,2$ and the celestial amplitude is 
\begin{equation}
\label{eq:l1bbampcel}
\tilde{\mathcal{A}}_{2;E}^{E}=-4g_e\pi  \int d^2 x_\perp ~\frac{\Gamma(\Delta_1+\Delta_2)}{(-i \hat{p}^\perp_{12}\cdot \vec{x}_\perp)^{\Delta_1+\Delta_2}}a^E(\vec{x}_\perp)~.
\end{equation}
To match with the celestial four-point amplitude we require the background potential to be
\begin{equation}
a^E(\vec{x}_\perp)=(2\pi)^2i g_e \int d^2 \bar{x}_\perp G_\perp(\vec{x}_\perp, \vec{\bar{x}}_\perp) \frac{\Gamma(\Delta_3+\Delta_4)}{(-i \hat{p}^\perp_{34}\cdot \vec{\bar{x}}_\perp)^{\Delta_3+\Delta_4} }~
\end{equation}
which is the result of \cite{deGioia:2022fcn} in the spin-1 case.

We now turn to the four point amplitude of a electrically charged particle (momenta $p_1$ and  $p_2$) scattering from a magnetically charged particle (momenta $p_3$ and $p_4$). The tree-level result is 
\begin{align}
\mathcal{A}_4^{EM}=4(2\pi)^4i \frac{4\pi {g}_m g_e}{e^2 t} \frac{\epsilon(p_1,p_2,n,p_4)}{ n\cdot(p_1+p_2)} \delta(\sum p_i)
\end{align}
which gives the Mellin transformed amplitude
\begin{align}
\tilde{\mathcal{A}}_{4,eik}^{EM}=& \frac{-32\pi^3 i {g}_m g_e \epsilon(\hat{p}_1,\hat{p}_2,n,\hat{p}_4)}{e^2 n\cdot \hat{p}_{12}} \int d^2 x_\perp \int d^2 \bar{x}_\perp G_\perp(\vec{x}_\perp, \vec{\bar{x}}_\perp) \nn\ \\
& \kern+100pt \times\frac{\Gamma(\Delta_1+\Delta_2)}{(-i \hat{p}^\perp_{12}\cdot \vec{x}_\perp)^{\Delta_1+\Delta_2} }
\frac{\Gamma(\Delta_3+\Delta_4)}{(-i \hat{p}^\perp_{34}\cdot \vec{\bar{x}}_\perp)^{\Delta_3+\Delta_4} }~
\end{align}
where we use the notation $\hat{p}_i^\mu= {p}^\mu_i/\eta_i \omega_i$.
On the other hand, the two-particle electric scattering in a dual background
\begin{equation}
\bar{A}^M_\mu(p)=\pi \delta(p^-) \bar{a}^M(p_\perp) \epsilon_{\mu \kappa \nu \lambda} \frac{n^\kappa p^\nu q_s^\lambda}{n\cdot p} 
\end{equation}
is
\begin{equation}
\label{eq:l1bbampcmag}
\tilde{\mathcal{A}}_{2;M}^{E}=\frac{-2g_e\pi \epsilon(\hat{p}_1, n, \hat{p}_2, q_s)}{n\cdot \hat{p}_{12}} \int d^2 x_\perp ~\frac{\Gamma(\Delta_1+\Delta_2)}{(-i \hat{p}^\perp_{12}\cdot \vec{x}_\perp)^{\Delta_1+\Delta_2}}a^M(\vec{x}_\perp)~.
\end{equation}
This matches the eikonal amplitude if we identify $q_s=\hat{p}_4$ and 
\begin{align}
a^M(\vec{x}_\perp)=\frac{4\pi}{e^2 }\left. a^E(\vec{x}_\perp)\right|_{g_e\to -g_m}~.
\end{align}
We could of course take the opposite perspective and compute the scattering of a magnetic particle in an electric background in which case we find essentially the same result but we must identify $q_s$ with the momentum of the electrically charged particle i.e.  $\hat{p}_2$.

%%%%%%%%%%%%%%%%%%%%%%%%%%%%%%%%%%%%%%%%%%%%%%%%%%%%%%%%%%%
\section{
S-duality from
T-duality in boundary actions}
\label{sec:bdact}
%%%%%%%%%%%%%%%%%%%%%%%%%%%%%%%%%%%%%%%%%%%%%%%%%%%%%%%%%%%

Let us return to the issue of the boundary interpretation of a bulk S-duality transformation. 
S-duality in four-dimensional gauge theories, when dimensionally reduced to two dimensions, is known to become T-duality \cite{Bershadsky:1995vm,Harvey:1995tg}. Viewing a boundary description as a null reduction of the bulk theory, it is natural to conjecture that in the two-dimensional celestial boundary theory S-duality becomes T-duality. Proving this is beyond the scope of this work, but we can see that in the vacuum sector of the classical theory there is a natural connection via a Chern-Simons description of the boundary conditions. 

 We start by considering the boundary conditions of Maxwell theory at null infinity. Using retarded coordinates  $(r, u, z, \bar{z})$ where
\begin{equation}
ds^2=-du^2-2 du dr +2 r^2 \gamma_{z\bar z} dz d\bar{z}
\end{equation}
with $\gamma_{z\bar{z}}=2/(1+z \bar{z})^2$, future null infinity ($\mathscr{I}^+$) corresponds to $r\to \infty$. 
We consider a gauge potential with the asymptotic expansion
\begin{equation}
A_u=\sum_{n=2}^\infty \frac{A_u^{(n)}}{r^n}~,~~~A_r=\sum_{n=2}^\infty \frac{A_r^{(n)}}{r^n}~,~~~A_{z/\bar z}=\sum_{n=0}^\infty \frac{A_{z/\bar z}^{(n)}}{r^n}.
\end{equation}
In Lorenz gauge the components $A_{z/\bar z}^{(0)}$ provide the Cauchy data which can be used to determine the full gauge field, see \cite{Strominger:2017zoo}. There is a similar expansion on past null infinity ($\mathscr{I}^-$).
 We define a vacuum sector to comprise solutions with boundary conditions at null infinity 
\begin{equation}
\label{eq:bd_cond}
\left. F_{z\bar z}^{(0)}\right|_{\mathscr{I}^+}=\left. F_{uz}^{(0)}\right|_{\mathscr{I}^+}=\left. F_{u \bar z}^{(0)}\right|_{\mathscr{I}^+}=0
\end{equation}
where $F^{(0)}_{\mu\nu}$ is the leading, $\mathcal{O}(r^0)$, term in the large-$r$ expansion of the field strength near ${\mathscr{I}^+}$ and with analogous conditions at ${\mathscr{I}^-}$. In particular, this implies that the gauge potential is given by the Goldstone mode $\phi(z,\bar z)$ \cite{He:2014cra},
\begin{align}
A^{(0)}_{z/\bar z}&
=\partial_{z/\bar z}\phi(z, \bar{z})~.
\end{align}
 The S-duality map for $\theta=0$, \eqref{eq:duality_map}, implies that S-dual field strength satisfies 
 \cite{Strominger:2015bla}\\
\begin{equation}
W^{(0)}_{uz}=\frac{4\pi i}{e^2}F_{uz}^{(0)}~,~~~W^{(0)}_{u\bar z}=-\frac{4\pi i}{e^2}F_{u\bar{z}}^{(0)}
\end{equation}
 and so 
\begin{equation}
B_z^{(0)}=\frac{4\pi i}{e^2}A_{z}^{(0)}~,~~~B_{\bar{z}}^{(0)}=-\frac{4\pi i}{e^2}A_{\bar{z}}^{(0)}~.
\end{equation}
Introducing an S-dual Goldstone mode $\phi^M(z,\bar z)$,
\begin{equation}
    B_{z/\bar z}^{(0)}=\partial_{z/\bar z} \phi^M(z,\bar z),
\end{equation}
we have the duality relations
\begin{equation}
\label{eqn:Ltdual}
\partial_z \phi^M=\frac{4\pi i}{e^2} \partial_z \phi~,~~~\partial_{\bar{z}} \phi^M=-\frac{4\pi i}{e^2} \partial_{\bar{z}} \phi~
\end{equation}
which are essentially those of two-dimensional T-duality. 

We can explicitly see this for the conformal primary gauge potential \eqref{eq:cpw} and the magnetic dual potential \eqref{eq:dual_mom_cpw} with $\Delta=1$, which can be expressed in terms of scalar Goldstone modes,
\begin{equation}
    A^{\Delta=1}_{z,a}=\partial_z \alpha_{a}^{\Delta=1}, 
    \quad A^{M,\Delta=1}_{z,a}=\partial_z \alpha_{a}^{M,\Delta=1}.
\end{equation}
For positive helicity, $a=z_q$, the Goldstone mode and its dual are
\begin{equation}
\alpha_{z_q}^{1} =\frac{1}{z-z_q}, \quad \alpha_{z_q}^{M,1} =\frac{i}{z-z_q}.
\end{equation}
After accounting for the factor of $4\pi/e^2$ between \eqref{eq:AbarM} and \eqref{eq:dual_pot} this clearly obeys \eqref{eqn:Ltdual}. 

It is interesting to formulate the duality \eqref{eqn:Ltdual} in terms of an action for the Goldstone modes. Focusing on the boundary conditions \eqref{eq:bd_cond} on $\mathscr{I}^+$ we see that these are the three-dimensional equations of motion following from the Chern-Simons action
\begin{equation}
S_{bd}=
\frac{1}{\alpha} \int_{\mathscr{I}^+} d^3x ~\big[ A^{(0)}_u F_{z\bar{z}}^{(0)}-A^{(0)}_z F_{u\bar{z}}^{(0)}+A^{(0)}_{\bar{z}} F_{uz}^{(0)}\big]
\end{equation}
where $\alpha=e^2/4\pi$.
However, to properly define the variational problem we must take into account the contributions at the boundary, here $\mathscr{I}^+_\pm$ i.e. at $u=\pm \infty$\footnote{This is  well-known in the context of the quantum Hall effect, see \cite{Tong:2016kpv} for a review.}.
 In short, one must add additional chiral scalar degrees of freedom at the boundary which are described by the Floreanini-Jackiw action \cite{Floreanini:1987as}
\begin{equation}
S_{FJ}=\frac{1}{\alpha}\sum_{\sigma=\pm} \int_{\mathscr{I}^+_\sigma } d^2 z ~\Big[\partial_z \phi^\sigma \partial_{\bar{z}} \phi^\sigma-v^2 (\partial_z \phi^\sigma)^2\Big]
\end{equation}
where $v$ is the velocity of the edge modes which we take to be zero, $v=0$. Focusing on the $\mathscr{I}^+_+$ component of the boundary, and dropping the index on $\phi$, we recall the standard derivation of two-dimensional T-duality by re-writing the action as 
\begin{align}
S_{bd}^+=\int_{\mathscr{I}^+_+ } d^2 z ~\Big[\frac{1}{\alpha} C_z C_{\bar{z}} +  \phi^M (\partial_z C_{\bar{z}}-\partial_{\bar{z}}C_z)\Big]~.
\end{align}
Integrating out the Lagrange multiplier $\phi^M$ we have
\begin{align}
\partial_z C_{\bar{z}}-\partial_{\bar z} C_z=0\Rightarrow C_z =\partial_z \phi~,~~~\text{and}~~~C_{\bar{z}}=\partial_{\bar{z}}\phi
\end{align} 
so that we recover the original action. Alternatively, we can integrate out the variables $C_a$ by imposing the equation of motion 
\begin{align}
\frac{1}{\alpha} C_z=\partial_z \phi^M~, ~~~\frac{1}{\alpha}C_{\bar z}=-\partial_{\bar z} \phi^M
\end{align}
 where, as the action is Euclidean, we find Euclidean analogues of the T-dual relations \eqref{eqn:Ltdual}
giving rise to the dual action
\begin{align}
S_{bd}^+=\alpha \int_{\mathscr{I}^+_+ } d^2 z ~\Big[ \partial_z\phi^M \partial_{\bar{z}} \phi^M \Big].
\end{align}
This is the same as the original action but with the S-dual coupling $\tau= i/\alpha\to -1/\tau$. Thus we see the connection between S-duality and this T-duality type transformation.

In \cite{Kapec:2021eug}, see also \cite{Kapec:2022hih, He:2024ddb}, a pseudo-Goldstone action was constructed to reproduce the known soft factors and IR divergences of QED. This action is local when formulated in terms of the shadow edge mode but, as the authors point out and on which we focus, in two-dimensions the Goldstone mode is proportional to its shadow and so the Euclidean Goldstone mode soft action is already local. The action is essentially the same free boson action above but with a coupling, $\tau_{\text{eff}}=4\pi i/(e^2 \ln \Lambda/\mu)=i/\alpha_{\text{eff}}$ which incorporates the UV ($\Lambda$)  and IR  ($\mu$) cut-offs.  Here we include the logarithmic cut-off dependence in the effective dual coupling. This is somewhat reminiscent of the coupling in $\mathcal{N}=2$ super-symmetric gauge theory where S-duality has been extensively studied, for a review see for example \cite{Alvarez-Gaume:1996ohl}. Electrically charged particles, of momentum $p_j$ and charge $g_{ej}$, couple to the Goldstone boson via the current $\mathcal{J}^e_{z/\bar{z}}=\partial_{z/\bar{z}} \sum_j g_{ej} \log(-p_j \cdot \hat{q}(z))$. The action is 
\begin{align}
S_{\text{eff}}[\vartheta]=\frac{1}{\alpha_{\text{eff}}} \int d^2z~ \partial_z \vartheta \partial_{\bar{z}} \vartheta
	-\frac{i}{2\pi}\int d^2z~ (\partial_z \vartheta \mathcal{J}^{e}_{\bar{z}}+\partial_{\bar{z}} \vartheta \mathcal{J}^{e}_{z})~.
\end{align}
 Introducing a Lagrange multiplier $\vartheta^M$ and performing the two-dimensional T-duality transformation by integrating out the Goldstone field, we find 
\begin{align}
S_{\text{eff}}[\vartheta^M]=\alpha_{\text{eff}}\int d^2z ~ \partial_z \vartheta^M\partial_{\bar{z}} \vartheta^M-\frac{i}{2\pi} \int d^2z ~(\partial_z \vartheta^M \mathcal{J}^z_m+\partial_{\bar{z}} \vartheta^M \mathcal{J}^{\bar{z}}_m) -\frac{1}{2\alpha_{\text{eff}}}\int d^2 z \frac{\mathcal{J}^2_m}{(2 \pi)^2}
\end{align}
where 
\begin{align}
\mathcal{J}_m^{z/\bar{z}}=\pm \partial_{\bar{z}/z} \sum_j {g}^{\text{eff}}_{mj} \log(-p_j \cdot \hat{q}(z))
\end{align}
with ${g}^{\text{eff}}_{mj}=-g_{ej}\alpha_{\text{eff}}$. This action describes a dual boson coupling to magnetically charged particles with S-dual couplings $\tau_{\text{eff}}\to -1/\tau_{\text{eff}}$ and $g_{ej}\to -g^{\text{eff}}_{mj}$. There is a factor of $i$ in this expression compared to \eqref{eq:td1} and \eqref{eq:td2} which is related to the Euclidean signature of the effective action. A similar factor was found in the relation between the Euclidean effective action and the Lorentzian soft on-shell action \cite{ He:2024ddb}.

We see that the T-duality introduces the background coupling $\partial \vartheta^M \mathcal{J}_m$, used in the pseudo-Goldstone boson action \cite{Kapec:2021eug},  for the winding modes which reproduces the magnetic soft-theorem \cite{Strominger:2015bla}. It additionally produces a quadratic term, $\mathcal{J}_m^2$, but this is independent of the dual-Goldstone mode and simply ensures that the IR divergent term is independent of the description, electric or magnetic, that is used.

%%%%%%%%%%%%%%%%%%%%%%%%%%%%%%%%%%%%%%%%%%%%%%%%%%%%%%%%%%%
\section{Outlook}
\label{sec:outlook}
%%%%%%%%%%%%%%%%%%%%%%%%%%%%%%%%%%%%%%%%%%%%%%%%%%%%%%%%%%%

There are a number of directions suggested by the above results. It would be interesting to connect the form of the amplitudes in non-trivial backgrounds, particularly in the eikonal limit, with the pairwise little group \cite{Zwanziger:1972sx}, pairwise helicity \cite{Csaki:2020inw, Csaki:2022tvb} and pairwise celestial representations \cite{Lippstreu:2021avq}. While we did not consider asymptotic scattering states with different electric and magnetic charges, such a description could potentially be applied to the combined background and probe system, at least in the case where the background has a particle interpretation. Of course it is also natural to consider higher-point amplitudes where states with different charges would appear. 

At higher-points one could also study the behaviour of amplitudes in soft and collinear limits. Soft-theorems in gauge and gravity theories are known to be related to asymptotic symmetries and conservation laws \cite{Strominger:2017zoo}. The algebra of symmetries can be found by analysis of multi-soft and collinear limits. For minimally coupled gravity, the algebra has been found to be related to the $w_{1+\infty}$ algebra and in gauge theory to the s-algebra \cite{Strominger:2021mtt, Guevara:2021abz}. Generalisations and deformations of these algebras are known to occur in various contexts: non-minimal couplings \cite{Melton:2022fsf}, non-zero cosmological constant \cite{Taylor:2023ajd}, Moyal deformations \cite{Monteiro:2022xwq}, and loop corrections \cite{Costello:2022upu}, see also \cite{Ren:2022sws, Costello:2022wso, Bittleston:2022jeq, Bittleston:2024rqe}.
In \cite{Melton:2022fsf} it was shown that deformations of the gauge theory algebra are also generated by expanding around shockwave backgrounds. For self-dual radiative backgrounds, the leading terms in the tree-level OPEs were undeformed by the background \cite{Adamo:2023zeh}. Studying the algebra for more general backgrounds, including magnetic backgrounds,  would provide more information regarding their holographic interpretation.

A boundary realisation of electromagnetic duality was constructed in \cite{Freidel:2018fsk}
by adding additional boundary degrees of freedom, in particular for the dual potential. It would be interesting to understand the action of T-duality in this picture and if there is any relation to the duality invariant, doubled sigma models \cite{Tseytlin:1990nb, Tseytlin:1990va, Hull:2004in}. In \cite{Freidel:2018fsk}, see also \cite{Hosseinzadeh:2018dkh}, it was noted that the algebra of magnetic and electric soft-charges have a central extension and it would be particularly interesting to see if this can be found in celestial OPE through an analysis of the soft limits of scattering amplitudes. This is particularly interesting as a similar phenomenon was found in the algebra \cite{Akhoury:2022sfj, Akhoury:2022lkb} of supertranslations and their duals \cite{Kol:2019nkc, Godazgar:2019dkh}.

It would also be interesting to consider the generalisation to non-Abelian gauge theory. In this case, even the IR physics is more complicated though the IR divergences are still known to factorise, see \cite{Agarwal:2021ais} for a recent review. It has been recently shown that for any massless gauge theory the all-order IR divergences with a colour-dipole structure can be reproduced by a theory of two-dimensional Lie-algebra-valued free bosons \cite{Magnea:2021fvy}. In the non-Abelian case both the bulk S-duality and the boundary T-duality would be significantly richer and so would provide a more complete test of the map between these dualities.

%%%%%%%%%%%%%%%%%%%%%%%%%%%%%%%%%%%%%%%%%%%%%%%%%%%%%%%%%%%
\subsection*{Acknowledgements}
%%%%%%%%%%%%%%%%%%%%%%%%%%%%%%%%%%%%%%%%%%%%%%%%%%%%%%%%%%%
We would like to thank Riccardo Gonzo for collaboration at initial stages of this work. We would also like to thank Sangmin Choi, Riccardo Gonzo, Andy Strominger and Chris White for discussions and comments on draft versions of this work. The work of NM was supported by the Science and Technology Facilities Council (STFC) Consolidated Grants ST/T000686/1 “Amplitudes, Strings \& Duality” and ST/X00063X/1 “Amplitudes, Strings \& Duality”. No new data were generated or analysed during this study. AP is supported by the European Research Council (ERC) under the European Union’s Horizon 2020 research and innovation programme (grant agreement No 852386). This work was supported by the Simons Collaboration on Celestial Holography. 

\appendix

\section{Conventions}
\label{app:Conventions}

\begin{itemize}
	\item We work with mostly plus signature i.e. $(-+++)$ and for the totally anti-symmetric Levi-Civita tensor, $\epsilon_{\mu\nu\rho\sigma}$, take $\epsilon_{0123}=-1$.
	\item We use the parameterisation 
	\begin{align}
	p^\mu&=\eta \omega
	(1+|z|^2, z+\bar{z},-i(z-\bar{z}), \iota(1-|z|^2))
	\end{align}
	for massless external momenta with $\eta=1$ for incoming momenta, $-1$ for outgoing and $\iota^2=1$. 
\item	Our conventions for the Fourier transform are 
\begin{equation}
\Phi(x)=\int \frac{d^4p}{(2\pi)^4} e^{ip\cdot x}\bar \Phi(p)\,, \quad \bar \Phi(p)=\int d^4x~ e^{-ip\cdot x} \Phi(x)\,.
\end{equation}
\end{itemize}

\section{Momentum space log mode}
\label{app:log}
We compute the Fourier transform of the log mode \eqref{eq:log}
\begin{align}
\bar{{A}}^{\text{log}}_{\mu; a}=\bar{{V}}^{\text{log},\pm}_{\mu; a} +i p_\mu \bar{\alpha}^{\rm log}_a~,
\end{align}
with 
\begin{align}
\bar{{V}}^{\text{log}, \pm}_{\mu; a}=2(\pm i)\partial_a q^\nu \int d^4 x e^{-ip\cdot x} \int_0^\infty d\omega_q \frac{x_\mu x_\nu}{x^2}e^{\pm i \omega_q q\cdot x -\omega_q \epsilon}~.
\end{align}
Note that here we have defined incoming and outgoing solutions but we have not regularised the $1/x^2$. 
The result depends strongly on the location of the singularities. For example, let us consider the integral with the analogue of advanced/retarded prescription
\begin{align}
\int d^4 x   \frac{x_\mu x_\nu}{x^2\pm 2 i x^0 \epsilon +\epsilon^2}e^{-i p \cdot x}&=-(2\pi)^3\partial_{p_\mu} \partial_{p_\nu} \frac{\Theta(\pm p^0)}{2\mathbf{p}} \delta( p^0\mp\mathbf{p})\nn\\
&=(2\pi)^3\partial_{p_\mu} \partial_{p_\nu} \Theta(\pm p^0) \delta(p^2)
\end{align}
where $\mathbf{p}=|\vec{p}|$. Alternatively, if we choose the Feynman prescription, which is what we use in the main text, 
\begin{align}
\int d^4 x   \frac{x_\mu x_\nu}{x^2 -i \epsilon}e^{-i p \cdot x}&=-\partial_{p_\mu} \partial_{p_\nu} \frac{(2\pi)^2i}{p^2-i \epsilon }\nn\\
&=2(2\pi)^2i\left(- \frac{4p_\mu p_\nu}{p^6}+\frac{\eta_{\mu\nu}}{p^4}\right)~,
\end{align}
where in the last line we drop the $i\epsilon$ prescription for simplicity. 
We thus find 
\begin{align}
\bar{{V}}^{\text{log}, F, \pm}_{\mu, a}&=\mp 4(2\pi)^2 \int d\omega_q
\left(\pm \frac{ 4\omega_q q_\mu \partial_a q\cdot p}{(p^2\mp 2 \omega_q p\cdot q)^3}
+\frac{ \partial_a q_\mu}{(p^2\mp 2\omega_q p\cdot q)^2}
\right)+p_\mu v_a(p,q)\nn\\
&= 2(2\pi)^2 
\frac{p\cdot q \partial_a q_\mu - q_\mu \partial_a q\cdot p}{p^2 (p\cdot q)^2}
+p_\mu v_a(p,q)
\end{align}
where the last term with
\begin{align}
    v_a(p,q)=-16\pi^2 \frac{\partial_a q\cdot p}{(p\cdot q)p^4}~
\end{align}
corresponds to a total derivative term in position space which drops out in the amplitude. Note that the final result is independent of the choice of incoming/outgoing and so we may drop the $\pm$ superscript .

\section{Integral identity for shadow transform of correlator}\label{app:shadow}
To take the shadow transform of an operator in a three-point function in  we make use of the integral 
\begin{align}
%\frac{\Delta-1}{\pi}
&\int d^d z' \frac{1}{|z-z'|^{2(d-\Delta)}|z'-z_1|^{\Delta+\tilde{\Delta}}|z'-z_2|^{\Delta-\tilde{\Delta}}}=\nn\\
&\frac{\pi^{d/2} \Gamma(\Delta-d/2)\Gamma(d/2-
\left(\tfrac{\Delta-\tilde{\Delta}}{2}\right))
\Gamma(d/2-
\left(\tfrac{\Delta+\tilde{\Delta}}{2}\right))
}{\Gamma(d-\Delta)\Gamma(\tfrac{\Delta-\tilde{\Delta}}{2})\Gamma(
\tfrac{\Delta+\tilde{\Delta}}{2})|z_{1}-z_{2}|^{2\Delta-d}|z-z_1|^{d-\Delta+\tilde{\Delta}}|z-z_2|^{d-\Delta-\tilde{\Delta}}}
\end{align}
which can be found using the Schwinger parameterisation and the Symanzik method for conformal integrals \cite{Symanzik:1972wj}, see also \cite{Dolan:2000uw}.

%%%%%%%%%%%%%%%%%%%%%%%%%%%%%%%%%%%%%%%%%%%%%%%%%%%%%%%%%%%%%%%%%%%%%%
\section{Dyonic Fourier transform}\label{integralcomp}
%%%%%%%%%%%%%%%%%%%%%%%%%%%%%%%%%%%%%%%%%%%%%%%%%%%%%%%%%%%%%%%%%%%%%%
The Coulomb part of the gauge field is simply Fourier transformed using
\[
\int \hat{d}^{4} q  \frac{\hat{\delta}\left(2 q \cdot p\right)}{q^{2}}e^{iq\cdot K} = -\frac{1}{16\pi}\delta(p\cdot K)\log|K|^2,
\]
however the dual gauge field
\[
\tilde{A}^\mu &= 2e\int \hat{d}^{4} q \frac{\hat{\delta}^{(+)}\left(q^{2}-2 q \cdot p_{0}\right)}{q^{2}} \frac{\epsilon^{\mu\nu\rho\sigma}\eta_\nu q_{\rho}p_{\sigma}\cos(q\cdot x)}{q\cdot\eta}
\]
requires more care due to the extra non-localities and complicated integrand.
Using the ``symmetric'' string singularity prescription \cite{Milton:1976jq}, we can express this as
\[
\tilde{A}^\mu &= \lim_{\epsilon\rightarrow 0}e\int \hat{d}^{4} q \frac{\hat{\delta}^{(+)}\left(2 q \cdot p\right)}{q^{2}} \epsilon^{\mu\nu\rho\sigma}\eta_\nu q_{\rho}p_{\sigma}\cos(q\cdot x)\left[\frac{1}{q\cdot\eta + i\epsilon} + \frac{1}{q\cdot\eta - i\epsilon}\right].
\]
We can solve this using Schwinger parametrisation, where we write
\[
\tilde{A}^\mu &= -2\pi\lim_{\epsilon\rightarrow 0}ie\int \hat{d}^{4} q \int ds\int dt e^{-is q^2}e^{-s\epsilon}e^{iq\cdot x-itq\cdot p } \epsilon^{\mu\nu\rho\sigma}\eta_\nu p_{\sigma}\\
& \kern+60pt \times \frac{\pd}{\pd \eta^\rho}\left[\int_0^\infty \frac{d\lambda}{\lambda} e^{i\lambda(q\cdot\eta +i\epsilon)} - \int_{-\infty}^0 \frac{d\lambda}{\lambda} e^{i\lambda(q\cdot\eta -i\epsilon)}\right]
\]
We can perform the $q,t$ and $s$ integrations
\[
\int_0^\infty ds\int_0^\infty dt \int \hat{d}^4q e^{-isq^2}e^{-s\epsilon}e^{iq\cdot (x-\lambda\eta -tp)} &= \int_0^\infty ds\frac{i}{(4\pi)^2}\frac{e^{-s\epsilon}}{s^2}\int_0^\infty dt e^{-i\frac{(x-\lambda \eta-tp)^2}{4s}}\\
&= \int_0^\infty ds\frac{i}{8\pi}\frac{e^{-s\epsilon}}{s}e^{-i\frac{(x-\lambda \eta)^2}{4s}}\delta(p\cdot x)\\
&= \frac{i}{4\pi}\delta(p\cdot x)\log(|x-\lambda\eta|^2) + \cl{O}(\epsilon,\text{constants}).
\]
We are then left with 
\[
\tilde{A}^\mu &= -\lim_{\epsilon\rightarrow 0}\frac{g}{2} \delta(p\cdot x)\epsilon^{\mu\nu\rho\sigma}\eta_\nu p_{\sigma}\frac{\pd}{\pd \eta^\rho}\left[\int_0^\infty \frac{d\lambda}{\lambda} e^{-\lambda\epsilon}\log(|x-\lambda\eta|^2) - \int_{-\infty}^0 \frac{d\lambda}{\lambda} e^{-\lambda\epsilon}\log(|x
+\lambda\eta|^2)\right]\\
&= -\lim_{\epsilon\rightarrow 0}\frac{g}{2} \delta(p\cdot x)\epsilon^{\mu\nu\rho\sigma}\eta_\nu p_{\sigma}x_\rho\left[\int_0^\infty \frac{d\lambda}{|x-\lambda\eta|^2} e^{-\lambda\epsilon} - \int_{-\infty}^0 \frac{d\lambda}{|x+\lambda\eta|^2} e^{-\lambda\epsilon}\right]\\
&= g \delta(p\cdot x)\frac{\epsilon^{\mu\nu\rho\sigma}\eta_\nu p_{\sigma}x_\rho}{\sqrt{x^2\eta^2 - (x\cdot\eta)^2}}\tan^{-1}\left(\frac{x\cdot\eta}{\sqrt{x^2\eta^2 - (x\cdot\eta)^2}}\right)
\]

where we have used
\[
\lim_{\epsilon\rightarrow 0}\int _0^\infty d\lambda~ \frac{e^{-\epsilon\lambda}}{|a\pm\lambda b|^2} = \frac{\frac{\pi}{2} \mp \tan^{-1}\left[\frac{a\cdot b}{\sqrt{a^2b^2 - (a\cdot b)^2}}\right]}{\sqrt{a^2b^2 - (a\cdot b)^2}}.
\]
We note that this can be related to an alternative formulation by using the fact that
\[\label{arctanident}
\tan^{-1}(x) = \text{sgn}(x)\frac{\pi}{2} - \tan^{-1}(\frac{1}{x}), ~~~~~\text{sgn}(\frac{x}{y}) = \text{sgn}(x)\text{sgn}(y).
\]

\bibliographystyle{utphys}
\bibliography{CWE.bib}

\end{document}